\documentclass[preprints,article,accept,pdftex,moreauthors]{Definitions/mdpi} 

\firstpage{1} 
\pubvolume{1}
\issuenum{1}
\articlenumber{0}
\pubyear{2025}
\copyrightyear{2025}
\externaleditor{ }
\datereceived{ } 
\daterevised{ } 
\dateaccepted{ } 
\datepublished{ } 
\hreflink{doi.org/10.3390/universe11080258} 
\doinum{10.3390/universe11080258}
\usepackage{xcolor}

\usepackage[normalem]{ulem}

\Title{Role of Thermal Fluctuations in Nucleation of Three-Flavor Quark~Matter}

\TitleCitation{Role of Thermal Fluctuations in Nucleation of Three-Flavor Quark~Matter}
\usepackage{soul}

\Author{
Mirco Guerrini 
 $^{1,2,}$*\orcidA, 
Giuseppe Pagliara $^{1,2}$\orcidB, 
Andrea Lavagno $^{3,4}$\orcidC~and 
 Alessandro Drago $^{1,2}$\orcidD}

\AuthorNames{Mirco Guerrini, Giuseppe Pagliara, Andrea Lavagno and Alessandro Drago}

\AuthorCitation{Guerrini, M.; Pagliara, G.; Lavagno, A.; Drago, A}

\address{%
$^{1}$ \quad Department of Physics and Earth Science, University of Ferrara, 44122 Ferrara, Italy;  
\\
$^{2}$ \quad Istituto Nazionale di Fisica Nucleare (INFN), 
 Sezione di Ferrara, 44122 Ferrara, Italy\\
$^{3}$ \quad Department of Applied Science and Technology, Politecnico di Torino, 10129 Torino, Italy; \\
$^{4}$ \quad Istituto Nazionale di Fisica Nucleare (INFN), 
 Sezione di Torino, 10125 Torino, Italy}

\corres{Correspondence: mirco.guerrini@unife.it}

\abstract{
We present a framework that aims to investigate the role of thermal fluctuations in matter composition and color superconductivity in the nucleation of three-flavor deconfined quark matter in the typical conditions of high-energy astrophysical systems related to compact stars.
It is usually assumed that the flavor composition is locally fixed during the formation of the first seed of deconfined quark matter, since a weak interaction acts too slowly to re‐equilibrate flavors. However, the matter composition fluctuates around its average equilibrium values at the typical temperatures of high-energy astrophysical processes. 
Here, we extend our previous two‐flavor nucleation formalism to a three‐flavor case. We develop a thermodynamic framework incorporating finite‐size effects and thermal fluctuations in the local composition to compute the nucleation probability as the product of droplet formation and composition fluctuation rates. 
Moreover, we discuss the role of color superconductivity in nucleation, arguing that it can play a role only in systems larger than the typical coherence length of diquark pairs. We found that thermal fluctuations in the matter composition led to lowering the potential barrier between the metastable hadronic phase and the stable quark phase. Moreover, the formation of diquark pairs reduced the critical radius and thus the potential barrier in the low baryon density and temperature regime.
}

\keyword{dense QCD; nucleation; compact stars; strange matter} 

\begin{document}



\section{Introduction}
\label{sec:intro}
The possibility of forming quark matter in compact stars is usually modeled as a first-order phase transition which is triggered by the nucleation of the first drop of a quark phase in a hadronic medium. Many papers have addressed this issue by trying to estimate the probability of nucleation in several astrophysical conditions, namely for different values for the temperature and baryon density, different types of compositions (two-flavor or three-flavor quark matter and hadronic matter with or without hyperons), and different models for the dense matter equation of state~\cite{Olesen:1993ek,heiselberg1995quarkmatterdropletformation,Bombaci:2016xuj,Mintz:2009ay,Iida:1997ay}. Also, the~effect of color superconductivity on quark matter has been studied (see, for example,~\cite{Bombaci:2006cs}). Finding the threshold for the quark nucleation process (namely the condition for which the nucleation time is comparable to the dynamical time scale) is crucial for addressing the possibility of the occurrence of deconfinement in the formation of hybrid stars (HSs) {during a core-collapse supernova (CCSN) \citep{Fischer_2018},} in protoneutron stars (PNS), and in the merger of two compact stars (BNSM) (see, for example,~\cite{Blacker:2020nlq}). Moreover, in the case in which strange quark matter (SQM) is absolutely stable (and thus at~least some compact stars are actually strange quark stars), a~correct determination of the nucleation threshold is key for understanding under which conditions compact stars can convert into strange quark stars (QSs), leading to the scenario in which hadronic stars and strange quark stars coexist (the so-called two-family scenario~\cite{Drago:2013fsa,Drago:2015cea,Drago:2015dea}).

A key issue, when computing nucleation in multicomponent systems concerns the concentrations of the several species of particles in the metastable phase and in the new phase. In~the seminal paper~\cite{Olesen:1993ek}, it is argued that the energy fluctuation leading to the formation of a stable droplet of the quark phase occurs on a time scale which is to the order of the time scale of strong interaction, namely $\sim$$10^{-23}$ s. During~this tiny amount of time, there is no chance that the composition of matter can change due to the occurrence of weak interactions, which are much slower. 
Thus, the~flavor composition is frozen during the nucleation, and~the droplets of quark matter have the same concentrations of up, down, and strange quarks of the hadronic metastable phase. Since the flavor compositions of hadronic matter and quark matter in $\beta$ equilibrium are different (especially when considering strangeness), the~first droplet of quark matter would be in an out-of-equilibrium phase.
The effect of this assumption is  substantially increasing the baryon density at which quark matter appears with respect to the stationary situation of thermodynamical equilibrium, in which the two phases are in mechanical, chemical, and~thermal equilibrium. 
In our recent work~\cite{Guerrini:2024gzu}, we  revised the standard formalism of nucleation in multicomponent systems by introducing the possibility of thermal fluctuations in the composition of the different species of particles in the case of two-flavor quark matter. The~fluctuations we considered in that work basically corresponded to the density fluctuations within the grancanonical ensemble of statistical mechanics, which were in no way related to the action of weak interactions but are related instead to the fact that the droplet of quark matter was immersed in a large ``bath'' of hadronic matter that behaved as a source of energy and a source of particles for the quark matter droplet. As~a result, by~considering those kinds of fluctuations, the~nucleation process was much more efficient with respect to the case of a frozen composition. In~this paper, we will extend the formalism developed in~\cite{Guerrini:2024gzu} to the case of three-flavor quark matter. In~particular, we will consider hadronic equations of state (EOSs) which include hyperons (hadronic matter with finite net strangeness) and two types of strange quark matter equations of state: (1) unpaired quark matter and (2) color--flavor-locked (CFL) quark matter, one of the possible candidates for color superconductivity in compact stars~\cite{Alford:2001zr,Alford_2005,Alford:2007xm}. 

{Indeed, it has been suggested that color superconductivity may be key to compact star phenomenology, as~it helps stiffen the EOS at a high density while retaining the quark degrees of freedom, contributing to supporting high-mass configurations. This applies to both the one-family HSs~\cite{Gartlein:2023vif,Blaschke:2022egm,Ivanytskyi:2022qnw} and the two-family QSs~\cite{Bombaci:2020vgw}. 
This is relevant in the case in which the maximum compact star mass lies near $\sim$(2.2--
2.3)$\,M_{\odot}$ \cite{Rezzolla:2017aly} and~even more if the $\sim$$2.6\,M_{\odot}$ object in GW190814~\cite{Abbott_2020} is a compact star\endnote{{While there are robust analyses, based on the multi-messenger signals of GW170817, indicating that the maximum mass of NSs is significantly below $2.6 M_{\odot}$ (see, for example,~\cite{Nathanail:2021tay}), such results cannot be directly applied to the two-family scenario because in such a scenario, the merger remnant is a differentially rotating QS whose properties are rather different with respect to the case of nucleonic stars (see~\cite{Zhou:2021upu}).}} \cite{Bombaci:2020vgw}.}


Moreover, color-superconducting phases could set in only if the droplet of the quark phase is large enough to allow for the creation of a diquark pairing correlation, whose typical size is to the order of $\sim$$1/\Delta$ \cite{Alford:2007xm}. In~this respect, as~we will show, the~critical radius for the formation of a stable quark matter droplet could be strongly affected by the possibility of CFL~phase formation. 

The paper is organized as follows. In Section~\ref{sec:nucleation}, nucleation formalism considering thermal fluctuations in the hadronic composition will be presented in detail. In~Section~\ref{sec:fwork}, we will describe our framework, which aims to include the role of color superconductivity. Our results will be highlighted in Section~\ref{sec:results}. A~summary and the conclusions will be reported in Section~\ref{sec:conclusions}. Finally, in~Appendix \ref{sec:eos}, the~EOS models we use will be~presented.

We will use thenatural units $\hbar=c=1$ and $k_B=1$.

\section{Nucleation~Theory}
\label{sec:nucleation}

Let us consider a closed system with two stable equilibrium solutions: a pure metastable hadronic phase $H$ (corresponding to an energy local minimum) and a stable, pure, deconfined quark phase $Q$ (corresponding to the energy global minimum).
Let us initially prepare the system in the metastable hadronic phase $H$ at a temperature $T$. Due to thermal or quantum fluctuations, the~system continuously explores configurations around the local minimum (and thus the~thermodynamical quantities around their equilibrium average values). The~probability of a generic thermal fluctuation is 
 \cite{Landau}
\begin{equation}
   \mathcal{P}\propto e^{-W /T},
\end{equation}
where $W$ is the energy difference between the fluctuated and  equilibrium configurations, corresponding to the work needed to pass from one configuration to the other with a reversible transformation. The~higher the temperature, the~larger the probability of having large fluctuations with respect to the minimum. 
The available configurations depend on the kinds of reactions that can occur. For~example, if~only strong interactions occur, then the~global flavor composition cannot be modified. On~the other hand, if~weak reactions are also involved, then configurations with different flavor compositions with respect to the initial equilibrium can be~explored.

We are interested in studying the phase transition between the local and global minima in~particular, namely the deconfinement of quarks from hadrons in dense matter. Fluctuations can locally produce small droplets of the deconfined quark phase. The~most likely path in the configuration space to pass from the local to the global minima is the one that passes near the lowest intervening saddle point of the potential between phases $H$ and $Q$ \cite{Langer:1969bc,Langer_1973}. In~our picture, this saddle point describes a configuration in which the system is all in the metastable hadronic phase, except for a single critical droplet of the stable deconfined quark phase. Namely, the~saddle point is a local maximum of the potential as a function of the volume occupied by the quark droplet, thus forming a mechanically unstable equilibrium configuration. 
This arises from the competition between a negative bulk term, which scales with the volume of the new quark phase, and~a positive surface term related to the surface tension, which scales with its interfacial area.
The quantity $W=E_{sp}-E_H$ between the energy of the saddle point configuration $E_{sp}$ and the energy $E_H$ of the metastable $H$ state represents the height of the potential barrier (usually called ``activation energy'') that separates the two minima $H$ and $Q$. Thus, if~a fluctuation generates a droplet of the deconfined quark phase with a volume smaller than the critical one, then the~droplet is mechanically unstable and disappears. However, if~the generated droplet is as large as or larger than the critical one, then the~potential barrier is overcome, and~the whole system will relax in the new stable phase $Q$. The formation of a critical droplet is usually referred to as nucleation\endnote{In principle, nucleation can be either inhomogeneous or homogeneous. The~former is more common in nature and occurs when an impurity triggers the phase transition. However, we are interested here in homogeneous nucleation, where the phase transition is triggered by an intrinsic phenomenon, namely by a thermodynamical fluctuation.}, and it is the trigger for a first-order phase~transition.

\subsection{{Nucleation~Rate}}
The rate at which the potential barrier is overcome (number of critical-sized droplets created in a unit volume in unit time) by a thermal fluctuation is~\cite{Langer:1969bc,Langer_1973}
\begin{equation}
   \Gamma=\Gamma_0 \,e^{-W /T}, \label{eq:langerrate}
\end{equation}
where $\Gamma_0$ is a prefactor which, {in~\cite{Langer:1969bc,Langer_1973}, is computed as a product of a statistical prefactor $\Omega_0$ (which measures the phase space volume of the region around the saddle point) and a dynamical prefactor $\kappa$ (the exponential growth rate of the droplet at the saddle point):
\begin{equation}
    \Gamma_0 = \frac{\kappa}{2\pi}\Omega_0.
\end{equation}

\noindent The 
 quantities $\kappa$ and $\Omega_0$ were explicitly computed in~\cite{Csernai_1992_prefactors,Venugopalan_1994}, where the homogeneous thermal nucleation theory~\cite{Langer:1969bc,Langer_1973} was extended for first-order phase transitions occurring in relativistic systems. The~statistical prefactor is
\begin{equation}
    \Omega_0=\frac{2}{3\sqrt{3}}\left(\frac{\sigma}{T}\right)^{3/2}\left(\frac{R_*}{\xi_q}\right)^4,
\end{equation}
where $\xi_q$ is the quark correlation length, $\sigma$ the surface tension, and $R_*$ the radius of the critical droplet (as will be discussed later in this section). 
For the dynamical prefactor, we will use the one computed in~\cite{Venugopalan_1994}, which incorporates the results of~
\cite{Langer:1969bc,Langer_1973,Csernai_1992_prefactors}
\begin{equation}
    \kappa = \frac{2 \sigma}{R_*^3 \left(\Delta w_*\right)}\left[\lambda T+2\left(\frac{4}{3}\eta + \zeta\right)\right],
\end{equation}
where $\Delta w_*$ is the enthalpy density difference between the two phases at the saddle point, $\lambda$ is the thermal conductivity, and $\eta$ and $\zeta$ are the shear and bulk viscosity of the hadronic matter, respectively.
The shear and bulk viscosity and the thermal conductivity were estimated in~\cite{Danielewicz:1984kt}. 
In this work, we will use the same values for the quantities playing a role in the prefactor as shown in~\cite{Bombaci:2009jt,Bombaci:2016xuj}. 
For a complete discussion on the prefactor, see~\cite{Kapusta:2007xjq}. However, the~exponential in Equation \eqref{eq:langerrate} dominates with respect to the prefactors. Therefore, in~many works (see for example,~\cite{Mintz:2009ay,Lugones:2015bya}), prefactors are replaced by simple dimensional-based expressions, such as~$T^4$, without~a relevant impact on the qualitative results~\cite{Carmo:2013fr}}.

Finally, the~nucleation time, namely the typical timescale after which a thermal nucleation event happens\endnote{{We emphasize that it is not the lifetime of a single droplet nor the time required for a single droplet to become critical;~rather, it is the time after which, statistically, a~single critical droplet is expected to appear in the system.}} is
\begin{equation}
    \tau=\frac{1}{V\,\Gamma},\label{eq:nucltime}
\end{equation}
where $\Gamma$ is the nucleation rate in Equation \eqref{eq:langerrate} and $V$ is the volume of the system. In~the context of compact stars, in~principle, one should compute the local nucleation rate at all points of the star where the hadronic phase is metastable and then integrate it over the volume. However, a~standard choice is to identify the total volume of the system as the innermost stellar region characterized by a sphere with a radius $\sim 100$ m, where thermodynamic quantities are almost constant (see, for example,~\cite{Bombaci:2016xuj,Mintz:2009ay}), assuming that the contribution of the rest of the star is negligible. Namely, $V\sim 10^{51}$ fm$^{3}$.  

Note that this formalism is general for the decay of a metastable state into a stable one and does not assume other specific features of the system (e.g., it does not assume that the droplet is a localized accumulation of particles, and it can also apply, for~example, to~the onset of superfluid phases by the formation of a sufficiently large vortex ring)  \cite{Langer:1969bc,Langer_1973}.

An alternative path for the nucleation process is quantum nucleation, in~which the potential barrier is overcome through quantum tunneling.
It has been shown that at low temperatures ($T\lesssim 10$ MeV), quantum nucleation becomes more efficient than thermal nucleation~\cite{Bombaci:2016xuj,Guerrini:2024gzu}. An~estimate of the quantum nucleation rate can be found in~\cite{Iida:1997ay,Iida:1998pi}. However, in~this work, we will focus on thermal nucleation, which is more efficient for the typical temperatures of PNSs, CCSNe, and~BNSMs.

{As reported above, the~dominating term of the thermal nucleation rate in \mbox{Equation \eqref{eq:langerrate}} is the exponential, which depends on the temperature $T$ and the activation energy \mbox{$W=E_{sp}-E_H$.} The~goal in the remaining part of this section is thus the correct individuation of the energy of the system in the saddle point configuration $E_{sp}$ and the evaluation of~$W$.}
%
\subsection{{Flavor~Conservation}}
Let us begin by considering which configurations can be explored by the system through thermal fluctuations.
The typical timescale of a fluctuation in the volume of the {deconfined} quark matter phase within {a system in} a state $H$ is the typical time over which a mechanically unstable droplet is created, shrinks back to the metastable phase, and~disappears to restore the equilibrium of the system. The~timescale for generating a critical droplet is set by the time needed to reach the saddle point configuration by means of a thermal fluctuation. 
{Given that deconfinement demonstrates a strong interaction-mediated phase transition, the~fluctuation generating the critical droplet has the typical timescale of strong interaction ($\tau_{strong}\sim 10^{-24}$ s). Since a weak interaction of the typical timescale is many orders of magnitude longer than that of a strong interaction,  we can assume that only the strong interaction is ``turned on''. Thus, only the configurations of the system that fulfill the strong interaction conservation laws can be explored during the relevant timescale.}
{In particular}, the~conserved quantities will be the baryon number $B$, the~non-leptonic electromagnetic charge\endnote{This is the~electromagnetic charge of hadrons and quarks specifically. For~example, in~pure nucleonic matter, this number equals the proton number. This conserved number can be replaced with the isospin.} $C$, the~strangeness $S$, and the leptonic number $L$. Note that the first three numbers can be remapped into the three quark flavors $u$, $d$, and $s$, which are indeed conserved by strong~reactions. 

While this assumption is generally accepted in the literature on quark nucleation, it is debated whether $B$, $C$, $S$, and $L$ are locally or globally conserved. As~suggested in~\cite{Glendenning:1992vb,Muller:1997tm,Hempel:2009vp}, the~total free energy is minimized if the two phases can share the conserved numbers. Thus, number conservations are global (and not local) unless the reactions that exchange conserved numbers from one phase to the other are suppressed because of some microphysical mechanisms. These mechanisms could include charge screening effects caused by long-range forces, the~slowness of specific reactions, or~the suppression of diffusion processes.
A comprehensive microscopic treatment should consider all the rates of strong reactions responsible for forming the critical droplet and exchanging conserved numbers between the droplet and its surroundings. 
However, those calculations can be simulated within a thermodynamical approach in two limiting cases\endnote{In~\cite{Constantinou:2025wxj}, an intermediate approach is adopted in which the electromagnetic charge neutrality is fulfilled in part locally and in part globally. In~principle, the~same framework could be applied for a partial local and partial global flavor conservation.}: 
(1) local flavor conservation, where the exchange of conserved numbers between the droplet and the surrounding is suppressed, and
(2) global flavor conservation, where the conserved numbers are in strong chemical equilibrium between the droplet and the surroundings.
In the former (latter) limit, the~equilibration time for the exchange of strong conserved numbers between the quark droplet and the surroundings is much longer (shorter) than the timescale of formation of the critical~droplet.

In~\cite{Bombaci:2016xuj} (and the references therein),~local number conservation (frozen flavor composition) is assumed for quark nucleation. On~the other hand, in~some works addressing the effect of phase transitions in heavy-ion collisions,~global number conservation is assumed. For~example, in~\cite{Greiner:1987tg,Greiner:1991us,Lavagno:2022orw}, the strangeness is globally (and not locally) conserved, leading to the so-called ``strangeness distillation'' effect. The~same approach, but~with global isospin conservation, was used in~\cite{DiToro:2006bkw}. 

\subsection{{Thermal Fluctuations in the Hadronic Flavor~Composition}}
\label{sec:subsectionfluctfwork}
Another aspect, which was disregarded in~\cite{Bombaci:2016xuj}, concerns the thermal statistical fluctuation of the particle composition at a finite temperature, which could also impact the efficiency of nucleation, as~discussed in~\cite{Norsen:2002qw,DiToro:2006bkw,Mintz:2009ay}. 
In this work, we will use the approach presented in~\cite{Guerrini:2024gzu}. Let us assume that the conserved number $k$ cannot be exchanged with the surroundings (local conservation of a number $k$). Thus, the~critical droplet of deconfined quarks will have the same fraction of $k$ as the preexisting hadronic phase. However, at~a finite temperature $T$, the~composition $\{Y_k^H\}$ of the initial phase must be considered to be a bulk average value. Locally, the thermodynamic quantities fluctuate around such average values. As nucleation is a local phenomenon, the~first critical droplet of deconfined quark matter could be produced in a small subsystem of the initial hadronic system, in which the composition $\{Y_i^*\}$ is different with respect to the average values $\{Y_i^H\}$ and more favorable for~nucleation.

In this framework, the~total nucleation rate will be
\begin{equation}
  \Gamma = \Gamma_0 \, e^{-\frac{W_{1}}{T}}e^{-\frac{W_{2}}{T}} \label{eq:probfluc},
\end{equation}
where the second exponential is the probability of generating a critical droplet of a deconfined quark phase $Q_*$ within a specific fluctuated hadronic subsystem $H_*$ characterized by a composition $\{Y_i^*\}$, while the first exponential is the probability that the subsystem $H_*$ exists in the initial hadronic system~\cite{Guerrini:2024gzu}.

\subsection{{Energy of the System in the Relevant~Configurations}}

We now need to compute $W_{1}$ and $W_{2}$. 
Let us assume that the initial system has a volume $V$, it is in the metastable hadronic phase $H$ with~a temperature $T$, and~it is characterized by the numbers $\{N_i^H\}$, where $i=B,C,S,e,\nu_e$ label  the baryon number, the~non-leptonic electromagnetic number, the~strangeness\endnote{{Note that in this work, we are defining ``strangeness'' as the number of strange quarks. Thus, $Y_S^H=Y_{\Lambda}^H+Y_{\Sigma^+}^H+Y_{\Sigma^0}^H+Y_{\Sigma^-}^H+2Y_{\Xi^0}^H+2Y_{\Xi^-}^H$. Moreover, note that this value can be, in principle, larger than one.}} number, the~net electron number, and net electronic neutrinos number\endnote{Any linear combination of them that can univocally describe the hadronic phase can also be used. Note that we are at this stage assuming that the initial hadronic phase is in strong equilibrium, namely the composition in terms of hadrons $\{Y_h^H\}$ (where $h=p,n,\Lambda,\Sigma^+,\Sigma^0,\Sigma^-,\Xi^0,\Xi^-,\Delta^{++},\Delta^{+},\Delta^{0},\Delta^{-}$) is fixed by the three numbers $B$, $C$, and $S$ and the strong equilibrium relations $\mu_h^H=\mu_B^H+C_h \mu_C^H+S_h \mu_S^H$, where $C_h$ and $S_h$ are the $C$ and $S$ charges of hadron $h$, respectively (e.g., $\mu_p=\mu_B+\mu_C$, $\mu_n=\mu_B$, $\mu_{\Lambda}=\mu_B+\mu_S$ ). }, respectively. It is useful to introduce the number densities $n_i^H=N_i^H/V$ and~the number fractions $Y_i^H=N_i^H/N_B^H=n_i^H/n_B^H$. The~energy of the initial hadronic phase is
\begin{equation}
    E_H = E_H(N_B^H,\{Y_i^H\},V,T) = V \varepsilon_H(n_B^H,\{Y_i^H\},T).
\end{equation} 
where $\varepsilon_H$ is the energy density.
Let us now focus on the saddle point configuration, namely the one in which the system is all in the hadronic phase except for the presence of a single critical spherical droplet of the deconfined quark phase. The~two phases are separated by an infinitely small layer, and~the energy of the finite-size deconfined quark phase inside the droplet is approximated with a bulk term $E_{Q*}$ plus a surface term $E_{\sigma}$. 
We will label the quantities inside the droplet with $Q_*$ and those in the external hadronic surroundings with $\widetilde{H}$.
The energy of this saddle point configuration is 
\begin{align}
    E_{sp} &= E_{Q*}+E_{\widetilde{H}}\\
        &= E_{Q}(N_B^{Q*},\{Y_i^{Q*}\},V_{Q*},T_{Q*})+E_{\sigma}+E_{H}(N_B^{\widetilde{H}},\{Y_i^{\widetilde{H}}\},V_{\widetilde{H}},T_{\widetilde{H}})\\
        &= V_{Q*}\varepsilon_{Q}(n_B^{Q*},\{Y_i^{Q*}\},T_{Q*})+E_{\sigma}+V_{\widetilde{H}} \varepsilon_{H}(n_B^{\widetilde{H}},\{Y_i^{\widetilde{H}}\},T_{\widetilde{H}}).
\end{align}

\noindent In principle, such a critical droplet could be generated at any point in the system. At~a finite temperature, the~thermodynamical quantities of the system are not equal everywhere. The~composition $\{Y_i^H\}$ corresponds to an average global composition that could be locally different in different subsystems. Thus, the~critical droplet of quark matter can be generated in a subsystem in which the thermodynamical quantities fluctuate with respect to the average equilibrium ones. 
Let us consider a configuration in which the system is all in the hadronic phase and contains a subsystem $H_*$ characterized by the same baryon number of the critical quark droplet $N_B^{H*}=N_B^{Q*}$ surrounded by a hadronic phase labeled with $\bar{H}$. 

The energy of this configuration is
\begin{align}
    E_{fl} &= E_{H*}+E_{\bar{H}}\\
        &= E_{H}(N_B^{H*},\{Y_i^{H*}\},V_{H*},T_{H*})+E_{H}(N_B^{\bar{H}},\{Y_i^{\bar{H}}\},V_{\bar{H}},T_{\bar{H}})\\
        &= V_{H*}\varepsilon_{H}(n_B^{H*},\{Y_i^{H*}\},T_{H*})+V_{\bar{H}} \varepsilon_{H}(n_B^{\bar{H}},\{Y_i^{\bar{H}}\},T_{\bar{H}}).
\end{align}

\noindent The minimum thermodynamic work $W_2$ needed to generate a deconfined quark droplet $Q_*$ in a subsystem $H_*$, assuming a closed system in which the total volume $V$, numbers $\{N_i^H\}$, and entropy $S_H$ (since we are interested in the minimum thermodynamic work, namely the adiabatic one~\cite{Landau}) are conserved, is
\begin{align}
    W_2 &= E_{Q*}+E_{\widetilde{H}}-(E_{H*}+E_{\bar{H}})+\sigma {A_{Q*}},\label{eq:W2E1}
\end{align}
where $E_{\sigma}=\sigma{A}_{Q*}$, ${A}_{Q*}$ is the surface area of the $Q*$ droplet and $\sigma$ is the surface tension between the hadronic and quark phases. 
The surface tension plays a key role in the first-order phase transition by increasing or decreasing the work required to form a quark matter droplet.
{Its value should be computed self-consistently with the used EOS models, and~it is, in~principle, temperature- and density-dependent (see, for example,~\cite{Fraga:2018cvr,Bessa:2008nw,Carmo:2013fr,Grunfeld:2024ihq}). Different techniques for microscopic calculations have been employed in the literature, obtaining highly uncertain and model-dependent results ranging from a few to hundreds of MeV/fm$^2$}
 (see,  for example,~\cite{Alford:2001zr,Bessa:2008nw,Grunfeld:2024ihq,Mintz:2009ay,Heiselberg:1992dx,Iida:1997ay,Palhares:2010be,Lugones:2015bya,Lugones:2018qgu,Fraga:2018cvr,Schmitt:2020tac,Ju:2021hoy}.
{Since this work aims to address the role of thermal fluctuations in the composition of and color superconductivity in nucleation, we will use a constant surface tension as an independent parameter, as~performed, for example,~in~\cite{Bombaci:2016xuj}. 
In principle, the~energy of the deconfined quark droplet also has a curvature term other than the bulk and surface ones. This term depends linearly on the droplet radius and the curvature tension, which in~principle is density- and temperature-dependent~\cite{Lugones:2015bya,Carmo:2013fr,Grunfeld:2024ihq}. However, given the uncertainty in both the curvature and surface terms, we neglect the curvature energy in this work, incorporating all the finite-size terms of the droplet energy into the surface term. In~particular, the~effect of including a curvature tension can be effectively incorporated by choosing a larger surface tension as a parameter (see, for example,~\cite{Bombaci:2006cs}).}

With the same assumptions, the~minimum work $W_1$ for generating the subsystem $H_*$~is
\begin{align}
    W_1 &= E_{H*}+E_{\bar{H}}-E_H.\label{eq:W1E1}
\end{align}

\noindent The conservation relations are
\begin{align}
    V&= V_{Q*}+V_{\widetilde{H}} =V_{H*}+V_{\bar{H}} \label{eq:VVQsVHt}\\
    N_i^H&= N_i^{Q*}+N_i^{\widetilde{H}}=N_i^{H*}+N_i^{\bar{H}} \,\,\,\,\,\, \text{ $i=B,C,S,L$}  \label{eq:NiHNiQsNiHt}\\
   S_H&= S_{Q*}+S_{\widetilde{H}}=S_{H*}+S_{\bar{H}} \label{eq:SHSQsSHt}.
\end{align}

\noindent Note that we are choosing the leptonic number $N_L=N_e+N_{\nu_e}$ for the conserved number and, in~principle, not $N_e$ and $N_{\nu_e}$ separately. However, here we are dealing with systems that are charge-neutral (in the sense of electromagnetic charge), namely $N_C^H-N_e^H=0$. Since $N_C^H$ is conserved in the whole system, $N_e^H$ is also conserved. Thus, given the conservation of $N_L$ and the electromagnetic charge neutrality, $N_e^H$ and $N_{\nu_e}^H$ are actually both conserved separately in the whole~system.

\subsection{{Small Droplet Volume~Approximation}}
Let us assume now that the volume occupied by the deconfined quark droplet and the particle number in it are just small fractions with respect to the total ones, namely $V\sim V_{\bar{H}} \sim V_{\widetilde{H}}  \gg V_{Q*} \sim V_{H*}$ and $N_i \sim N_i^{\bar{H}} \sim N_i^{\widetilde{H}} \gg N_i^{Q*} \sim N_i^{H*}$. Indeed, the~typical critical droplet volume is $V_{Q*}\sim (100-1000)$ fm$^3$, while the total volume of the system is $V\sim 10^{51}$ fm$^3$. 
Within this assumption, processes involving the small droplet do not lead to any appreciable change in the intensive thermodynamical quantities ($\mu_i$, $T$, and $P$) of the external hadronic phase, which thus remains nearly equal to the ones in the initial state $H$  \cite{Landau}. In particular, the~surrounding hadronic phase behaves as an external thermal bath and an external reservoir of particles for the deconfined quark droplet $Q_*$ and for the local fluctuated subsystem $H_*$:
\begin{align}
    P_H &\simeq P_{\widetilde{H}} \simeq P_{\bar{H}} \label{eq:PHtPH}\\
     T &\simeq T_{\widetilde{H}}\simeq T_{\bar{H}}  \label{eq:THtTH}\\
    \mu_i^{H}&\simeq \mu_i^{\widetilde{H}}\simeq \mu_i^{\bar{H}} .\label{eq:muiHtmuiH}
\end{align}

\noindent Using $E_I=S_IT_I-P_IV_I+\sum_i N_i^I \mu_i^I$ (where $I=Q_{*},H_{*},\widetilde{H},\bar{H},H$), by~substituting the conservation relations (Equations (\ref{eq:VVQsVHt}--\ref{eq:SHSQsSHt})) into Equations \eqref{eq:W2E1} and  \eqref{eq:W1E1} and~assuming the equalities in Equations~(\ref{eq:PHtPH}--\ref{eq:muiHtmuiH}), we obtain
\begin{align}
    W_2&= S_{Q*}(T_{Q*}-T)+ S_{H*}(T-T_{H*})-V_{Q*}(P_{Q*}-P_H)-V_{H*}(P_H-P_{H*})+\\&+\sum_iN_i^{Q*}(\mu_i^{Q*}-\mu_i^H)+\sum_iN_i^{H*}(\mu_i^H-\mu_i^{H*}) +\sigma {A}_{Q*}\label{eq:W1EsmallV} 
\end{align}
and
\begin{align}
    W_1 &= S_{H*}(T_{H*}-T)-V_{H*} (P_{H*}-P_H)+\sum_i N_i^{H*}(\mu_i^{H*}-\mu_i^{H}) \label{eq:W1EsmallV}.
\end{align}

\noindent Finally, the~total work is
\begin{align}
    W&=W_1+W_2 = E_{Q*}+E_{\widetilde{H}}-E_H +\sigma {A}_{Q*} =\\
    &= S_{Q*}(T_{Q*}-T)-V_{Q*}(P_{Q*}-P_H)+\sum_i N_i^{Q*}( \mu_i^{Q*}- \mu_i^{H})+\sigma {A}_{Q*}.\label{eq:WW1W2}
\end{align}

Since we are assuming that the quark phase is enclosed in a spherical droplet, we can write $V_{Q*}=4/3\pi R_*^3$ and ${A}_{Q*}=4\pi R_*^2$, where $R_*$ is the radius of the critical droplet (the droplet radius corresponding to the saddle point configuration). 

At this point, we want to fix all the free thermodynamic variables other than the ones that describe the initial conditions of the hadronic system.
\subsection{{Local Flavor Conservation in the Fluctuation Generating the Quark~Droplet}}
In our framework, the~flavor composition of the quark droplet $Q_*$ is equal to that of the fluctuated subsystem $H_*$ (see Section~\ref{sec:subsectionfluctfwork}):
\begin{align}
    Y_C^{Q*}& \equiv \frac{2}{3}Y_u^{Q*}-\frac{1}{3}Y_d^{Q*}-\frac{1}{3}Y_s^{Q*} = Y_C^{H*} \equiv Y_C^* \label{eq:localYC}\\
    Y_S^{Q*}&\equiv Y_s^{Q*} = Y_S^{H*}\equiv Y_S^*.\label{eq:localYS}
\end{align}

\noindent This choice corresponds to local number conservation~\cite{Hempel:2009vp} during quark droplet formation. We are thus assuming that while strong reactions lead to the generation of a $Q_*$ droplet in the $H_*$ subsystem, strong reactions and diffusion leading to the exchange of $C$ and $S$ between the droplet and the surroundings are suppressed (i.e., the~quasiparticles $S$ and $C$ {are exchanged on a time scale much larger than the typical time scale for  formation of the critical droplet.)}
Due to the construction, $N_B^{Q*}=N_B^{H*}\equiv N_B^*$, and~thus $N_i^{Q*}=N_i^{H*}\equiv N_i^*$, with~$i=C,S$.
In principle, $Y_C^*$ and $Y_S^*$ are, at~this stage, independent variables. One could remap these variables with a variation with respect to the equilibrium bulk values $Y_i^*=Y_i^H+\Delta Y_i$ (where $i=C,S$). If~$\Delta Y_i=0$, then we are back in the standard scenario in which the thermodynamic quantities of the system do not fluctuate and are equal everywhere (see, for example,~\cite{Bombaci:2016xuj}).

\subsection{{Electromagnetic Charge~Neutrality}}
Since we are interested in astrophysical applications, electromagnetic charge neutrality must be imposed. The~electromagnetic charge neutrality can be, in~principle, locally or globally fulfilled. Which of the two approaches better describes the mixed phase depends on the interplay between the surface tension and the Debye screening length~\cite{Heiselberg:1992dx,Voskresensky_2003,Constantinou:2025wxj}. Moreover, a~complete discussion should consider the contribution of electrostatic energy and compute how leptons are distributed to screen the droplet charge~\cite{Iida:1997ay,Iida:1998pi}. However, in~this work, we will assume, for~simplicity, local charge neutrality:
\begin{align}
    Y_e^{Q*}&=Y_C^{Q*}\equiv Y_C^* \label{eq:YeQYCQcn}\\
    Y_e^{H*}&=Y_C^{H*}\equiv Y_C^*\label{eq:YeHYCHcn}.
\end{align}

{In principle, fluctuations in the hadronic phase leading to an electrically charged subsystem
($Y_e^{H*}\neq Y_C^{H*}$) should also be considered. Notice  that the electric charge of the hadronic component is conserved during  generation of the deconfined quark droplet ($Y_C^{H*}=Y_C^{Q*}$). Therefore, a~fluctuation with $Y_C^{H*}-Y_e^{H*}\neq0$ implies an extra electrostatic energy in both $H_*$ and $Q_*$.
A more comprehensive investigation of the contribution of electrostatic energy and charge neutrality will be provided in future work.}

\subsection{{Assumptions for the Hadronic~Fluctuation}}
Moreover, let us assume that the fluctuation in the hadronic system leading to the subsystem $H_*$ occurs in mechanical and thermal equilibrium:
\begin{align}
    T_{H*}&=T_{\bar{H}}=T \label{eq:THeq}\\
    P_{H*}&=P_{\bar{H}}=P_H.\label{eq:PHeq}
\end{align}

\noindent We are thus assuming that in the fluctuation of the hadronic system, mechanical and thermal equilibrium are restored on a negligible timescale, while chemical equilibrium is reached on a longer timescale (see, for example,~\cite{Landau}). Thus, the~subsystem $H_*$ will have the same pressure and temperature as the surroundings but, in~general, a~different composition $\{Y_i^{H*}\}\neq\{Y_i^{H}\}$.

\subsection{{Role of~Neutrinos}}
Finally, neutrinos will be considered globally conserved~\cite{Hempel:2009vp} in the system (e.g., in the cases in which they are trapped inside the NS, namely during the first stage of the evolution of a PNS)
\begin{equation}
    \mu_{\nu_e}^{H}=\mu_{\nu_e}^{H*}=\mu_{\nu_e}^{Q*} \label{eq:neutrinosglobal}
\end{equation}
or untrapped once created (e.g., when the star cooled down, the~neutrinos' mean free path becomes larger than the star's size and they are thus free-streaming):
\begin{equation}
    \mu_{\nu_e}^H=\mu_{\nu_e}^{H*}=\mu_{\nu_e}^{Q*}=0  \label{eq:neutrinosuntrapped}
\end{equation}

\noindent In the latter case, the~lepton number is no longer conserved.
As we will see (see also~\cite{Hempel:2009vp}), in~both cases, neutrinos can be handled as an independent contributions to~EOSs.

Using all these conditions, the~works $W_1$, $W_2$, and $W$ become
\begin{align}
    W_1&= \sum_{i=B,C,S,e}N_i^{*}(\mu_i^{H*}-\mu_i^{H})\\
    &=\frac{4}{3}\pi R_*^3n_B^{Q*}\sum_{i=B,C,S,e} Y_i^*(\mu_i^{H*}-\mu_i^{H})
    \label{eq:W1localHeq}
\end{align}
\begin{align}
    W_2&= S_{Q*}(T_{Q*}-T)-V_{Q*}(P_{Q*}-P_H)+\sum_{i=B,C,S,e}N_i^{*}(\mu_i^{Q*}-\mu_i^{H*}) +\sigma {A}_{Q*}\label{eq:W2localHeqfull}\\
    &= \frac{4}{3}\pi R_*^3 \left[s_{Q*}(T_{Q*}-T)-(P_{Q*}-P_H)+n_B^{Q*}\sum_{i=B,C,S,e} Y_i^*(\mu_i^{Q*}-\mu_i^{H*})\right] +4\pi\sigma R_*^2
    \label{eq:W2localHeq}
\end{align}
\begin{align}
    W &= W_1+W_2\\
    &= S_{Q*}(T_{Q*}-T)-V_{Q*}(P_{Q*}-P_H)+\sum_{i=B,C,S,e}N_i^{*}(\mu_i^{Q*}-\mu_i^{H}) +\sigma {A}_{Q*}\\
    &= \frac{4}{3}\pi R_*^3 \left[s_{Q*}(T_{Q*}-T)-(P_{Q*}-P_H)+n_B^{Q*}\sum_{i=B,C,S,e} Y_i^*(\mu_i^{Q*}-\mu_i^{H})\right] +4\pi\sigma R_*^2 \label{eq:WlocalHeq},
\end{align}
where $s_{Q*}$ is the entropy~density.

For a specific EOS for the quark matter, $P_{Q*}$, $s_{Q*}$, and $\mu_i^{Q*}$ can be computed as a function of $(n_B^{Q*},\{Y_i^{Q*}\},T_{Q*})$, while $\mu_i^{H*}$ can be computed as a function of $(n_B^{H*},\{Y_i^{H*}\},T_{H*})$\endnote{Note that while $N_i^{H*}=N_i^{Q*}\equiv N_i^*$ and $Y_i^{H*}=Y_i^{Q*}\equiv Y_i^*$, we still have $n_B^{H*}\neq n_B^{Q*}$, since in general $V_{H*}\neq V_{Q*}$.}. 

\subsection{{Computing $n_B^{Q*}$, $T$, and $R_*$}}

\textls[-15]{Using the local $C$ and $S$ conservation in the droplet formation (\mbox{Equations (\ref{eq:localYC}) and (\ref{eq:localYS})}), the~electromagnetic charge neutrality (Equations (\ref{eq:YeQYCQcn}) and (\ref{eq:YeHYCHcn})), and the~mechanical  \mbox{(Equation~(\ref{eq:PHeq}))}} and thermal  (Equation (\ref{eq:THeq})) equilibrium of the hadronic fluctuation and the trapped (or untrapped) neutrinos condition (Equation (\ref{eq:neutrinosglobal}) or  Equation (\ref{eq:neutrinosuntrapped})), we are left with the independent variables $n_B^{Q*}$, $R_*$, $\{Y_i^*\}$ (or equivalently $\{\Delta Y_i\}$), and $T_{Q*}$, other than the input variables of the initial system $n_B^H,\{Y_i^H\},T$\endnote{Even if we are in general considering $\{Y_i^H\}$ as the input here, they are sometimes fixed by $n_B^H$, ($Y_L^H$) and $T$, such as if the initial system is in $\beta$ equilibrium without (with) trapped neutrinos. On~the other hand, $Y_C^H$ (e.g., in CCSNe~\cite{Oertel_2017}) or $Y_S^H$ (e.g., in heavy-ion collisions) remain independent variables if the dynamical timescale is shorter than the equilibration timescale of leptonic weak interaction and non-leptonic weak interaction, respectively (see~\cite{Hempel:2009vp,Oertel_2017}).}.

For the moment, let us keep $\{Y_i^*\}$ as independent variables. The~goal is thus to  somehow fix the remaining non-input independent variables $n_B^{Q*}$, $R_*$, and~$T_{Q*}$.

Usually (see, for example,~\cite{Landau,Bombaci:2006cs,Langer:1969bc,Mintz:2009ay}), it is assumed that the droplet of deconfined quarks is in thermal equilibrium with the surroundings:
\begin{equation}
    T_{Q*}=\widetilde{T}=T \label{eq:TQsTtT}
\end{equation}

Specifically, thermal equilibrium is reached in a negligible timescale during formation of the critical droplet. One can easily note that, under~this assumption, the~minimum work required to jump from one configuration to another is the difference between the free energies $F=E-ST$ of the two~configurations.

Let us explore some possibilities for fixing the other two variables.

\subsubsection{Small Degrees of~Metastability}
In~\cite{Guerrini:2024gzu}, we used the ``small degree of metastability''~\cite{Landau} approximation as shown in~\cite{Bombaci:2016xuj}. 
The system has a ``low degree of metastability'' if
\begin{eqnarray}
    \delta P_{H} = |P_{H}-P_x |&\ll& P_x\\
    \delta P_{Q^*} = |P_{Q^*}-P_x| &\ll& P_x,
\end{eqnarray}
where $P_x$ is the pressure at equilibrium when $R_* \rightarrow +\infty$ (plane surface) or $\sigma \rightarrow 0$ such~that
\begin{equation}
    \mu^{H}_k(P_x,T) = \mu^{Q}_k(P_x,T), 
\end{equation}
where $k$ labels every globally conserved charge and $\mu_k$ is the associated chemical potential.
Namely, $P_x$ is the equilibrium pressure of the two phases in a first-order phase transition at a temperature $T$ in bulk. 
In other words, a~system has a low degree of metastability if the overpressure needed in the metastable phase to balance the finite-size effects due to the surface tension is relatively~small.

Thus, within~this approximation, we can set the equation
\begin{equation}
   P_{Q*}\simeq P_{H} \label{eq:PQsmalldegrees}
\end{equation}
and substitute it as an equality together with Equation \eqref{eq:TQsTtT} in Equations \eqref{eq:W2localHeq} and \eqref{eq:WlocalHeq} to~obtain
\begin{equation}
    W_2=\frac{4}{3}\pi R_*^3 n_B^{Q*}\left[\sum_{i=B,C,S,e}Y_i^{*}\left(\mu_i^{Q*}-\mu_i^{H*}\right)\right]+4\pi \sigma  R_*^2,
\end{equation}
\begin{equation}
    W = \frac{4}{3}\pi R_*^3 n_B^{Q*}\left[\sum_{i=B,C,S,e}Y_i^{*}\left(\mu_i^{Q*}-\mu_i^{H}\right)\right]+4\pi \sigma  R_*^2,
\end{equation}
where $n_B^{Q*}$ is fixed using the equality in Equation \eqref{eq:PQsmalldegrees}.

Note that $\sum_iY_i^{*}\left(\mu_i^{Q*}-\mu_i^{H*}\right)$ is equal to $\mu_{Q*}-\mu_{H*}$, where $\mu_{I}=\sum_iY_i^I\mu_i^I=(P_I+\varepsilon_I-s_I T_I)/n_B^I$ is the Gibbs energy per baryon in the phase $I$ but~computed without neutrino contributions. Moreover, in~the absence of hadronic composition fluctuations ($Y_i^*=Y_i^H$ for $i=C,S$), we have $\sum_iY_i^{*}\left(\mu_i^{Q*}-\mu_i^{H}\right)$=$\mu_{Q*}-\mu_{H}$ again, computed without~neutrinos.

Let us now focus on fixing $R_*$. In~\cite{Guerrini:2024gzu}, we computed it as the droplet radius that maximizes $W_2$. However, this approach actually underestimates $R_*$. 
Indeed, $R_*$ must correspond to a mechanically unstable configuration such that once it is overcome by $\delta R$, the~whole system is free to reach the absolute minimum of the potential. However, a~droplet with a radius $R_2$ (such that $\max{W_2}=W_2(R_2)$) would not have this feature. 
Once a droplet with a radius $R_2$ is formed, a~subsystem $H_*$ characterized by a number of baryons $N_B^{*}=\frac{4}{3}\pi R_2^3 n_B^{Q*}$ and a composition $\{Y_i^*\}$ is converted into a deconfined quark $Q_*$ droplet. However, the~surrounding area is in the $\widetilde{H}\simeq H$ configuration and~not  $H_*$. Thus, it is not necessarily energetically convenient for the quark droplet to increase its radius by $\delta R$ and absorb other baryons. Indeed, the~$Q_*$ phase is energetically favorable with respect to $H_*$ but~not with respect to $H$. Namely, $\mu_{Q*}<\mu_{H*}$, but $\mu_{Q*}>\mu_{H}$ at fixed $P$ and $T$ values. 

One can easily note that the requested feature for the critical radius is fulfilled by a value of $R_*$ such that $\max{W}=W(R_*)$. A~configuration having a deconfined droplet with a radius $R_*$ is in a phase $Q_*$ that, even when adding the surface tension term, is more stable than the surrounding matter $\widetilde{H}\simeq H$.

Thus, $R_{*}$ corresponds to the maximum of $W$ (and not of $W_2$) such that
\begin{equation}
    R_*=\frac{2 \sigma}{n_B^{Q*}\left[\sum_{i=B,C,S,e}Y_i^{*}\left(\mu_i^{H}-\mu_i^{Q*}\right)\right]} \label{eq:Rcmetastab}
\end{equation}
and it exists only if $\sum_{i=B,C,S,e}Y_i^{*}\left(\mu_i^{H}-\mu_i^{Q*}\right)>0$. The~total work $W$ is thus
\begin{equation}
    W = \frac{16{\pi}}{3}\frac{\sigma^3}{\left[n_B^{Q*}\sum_{i=B,C,S,e}Y_i^{*}\left(\mu_i^{H}-\mu_i^{Q*}\right)\right]^2}.
\end{equation}

\subsubsection{Saddle-Point~Approach}
This approach states that $n_B^{Q*}$ is not fixed by any physical condition, and~thus it can have, in~principle, any value, indicating how big (in terms of the baryon number $N_B^*=n_B^{Q*}V_{Q*}$) the fluctuation is. 
In this subsection, we will still consider $T_{Q*}$ a free variable, as~the condition $T_{Q*}=T$ will naturally emerge in the~calculations.

In the spirit of the saddle point approach (where the most likely path in the configuration space to pass from the local to the global minima is the one that passes near the lowest intervening saddle point of the potential~\cite{Langer:1969bc,Langer_1973}), we will choose the $R_*$ value that maximizes $W$ (unstable mechanical equilibrium) and the $n_B^{Q*}$ and $T_{Q*}$ values that minimize it. 
The reason for this is that since $n_B^{Q*}$ and $T_{Q*}$ are, in principle, free, one should compute the total probability of nucleating as an integral over all the possible $n_B^{Q*}$ and $T_{Q*}$ values, but as the probability of a negative exponential of $W/T$, the~contributions related to the $n_B^{Q*}$ and $T_{Q*}$ values that minimize $W$ will~dominate. 

By minimizing $W$ (Equation (\ref{eq:WlocalHeq})) with respect to $T_{Q*}$, we obtain the thermal equilibrium condition
\begin{equation}
    T_{Q*}=T, \label{eq:tofindTQs}
\end{equation}

While minimizing with respect to $n_B^{Q*}$ and adding the condition $T_{Q*}=T$ we previously found, we obtain
\begin{equation}
    \sum_{i=B,C,S,e} Y_i^* \mu_i^H = \sum_{i=B,C,S,e} Y_i^* \mu_i^{Q*} \label{eq:tofindnBQs}
\end{equation}

\noindent Thanks to those conditions, $T_{Q*},n_B^{Q*}$ can be fixed, and all the thermodynamic quantities can be computed.
The $R_*$ value that maximizes $W$ is
\begin{equation}
    R_*=\frac{2 \sigma}{P_{Q*}-P_H},\label{eq:critRsaddle}
\end{equation}
which does indeed correspond to an (unstable) mechanical equilibrium~configuration.

The work $W_1$ can be computed using Equation \eqref{eq:W1localHeq}, while $W_2$ becomes
\begin{equation}
    W_2= -\frac{4}{3}\pi R^3 \left[P_{Q*}-P_H-n_B^{Q*}\sum_{i=B,C,S,e}Y_i^*\left(\mu_i^H-\mu_i^{H*}\right)\right]+4\pi \sigma R_*^2
\end{equation}
and the total work $W$ becomes
\begin{equation}
    W= -\frac{4}{3}\pi R_*^3 \left[P_{Q*}-P_H\right]+4\pi \sigma R_*^2\label{eq:critWsaddle}.
\end{equation}

\noindent By substituting Equation \eqref{eq:critRsaddle} into Equation \eqref{eq:critWsaddle}, we obtain
\begin{equation}
    W= \frac{16}{3}\pi \frac{\sigma^3}{(P_{Q*}-P_H)^2}.\label{eq:critWcsaddle}
\end{equation}

The resulting $W$ is the one used in~\cite{Mintz:2009ay}.

\subsection{{Computing $\{Y_i^*\}$}}
To conclude this section, let us come back to the choice of $\{Y_i^*\}$.
A certain subsystem of the system has a certain probability $\mathcal{P}_1$ to have the composition $\{Y_i^*\}$, and a deconfined quark droplet has a probability $\mathcal{P}_2$ to be generated in such a subsystem. Thus, complete calculations would need to compute the total nucleation probability by integrating over all of the possible fluctuated composition $\{Y_i^*\}$. However, since the nucleation probability is a negative exponential of the work $W$, the~composition associated with the lower total work $W$ will dominate the total probability. A~good choice for $\{Y_i^*\}$ would thus be the one that minimizes $W$. In~\cite{Guerrini:2024gzu}, we assumed that such a composition is equal to the composition of a quark phase in chemical equilibrium. In particular, we considered the fluctuation in the hadronic composition leading to a hadronic subsystem to have the same composition as a quark phase in $\beta$ equilibrium $Y_i^*=Y_i^{Q_{\beta}}$. Let us explicitly minimize W:
\vspace{3pt}
\begin{align}
    0=\left.\frac{\partial W}{\partial Y_C^*}\right|_{n_B^{Q*},Y_{S*},T}&=\frac{\partial}{\partial Y_C^*}\left[f_{Q*}+P_H-n_B^{Q*}\sum_{i=C,S,e}Y_i^* \mu_i^{H}\right]\\
    &= n_B^{Q*}\left(\mu_C^{Q*}+\mu_e^{Q*}-\mu_C^H-\mu_e^H\right)
\end{align}
where we have used Equation \eqref{eq:YeQYCQcn} as a constraint. Thus, the condition that minimizes $W$ with respect to $Y_C^*$ is
\begin{equation}
    \mu_C^{Q*}+\mu_e^{Q*}=\mu_C^H+\mu_e^H.\label{eq:muCmue}
\end{equation}

\noindent Similarly, the condition that minimizes $W$ with respect to $Y_S^*$ is
\begin{equation}
    \mu_S^{Q*}=\mu_S^H.\label{eq:muS}
\end{equation}

\noindent Note that by applying these conditions to Equation \eqref{eq:tofindnBQs} together with Equations~(\ref{eq:localYC}--\ref{eq:YeHYCHcn}), we~obtain
\begin{equation}
    \mu_B^{Q*}=\mu_B^H.\label{eq:muB}
\end{equation}

One can note that Equations~(\ref{eq:muCmue}--\ref{eq:muB}) correspond to the chemical equilibrium conditions with respect to strong interactions (see, for example,~\cite{Hempel:2009vp}). {Thus, even if we started with a different picture, our approach is analytically identical to what we would obtain when assuming a global (and not local) conservation of the charges conserved by the strong interaction and the chemical equilibrium with respect to them}.

Finally, let us consider what happens when the initial hadronic phase is in $\beta$ equilibrium with  trapped (untrapped) neutrinos. At~given $n_B^H,Y_L^H,T$ ($n_B^H,T$) values, the~$\beta$ equilibrium composition $\{Y_i^{H_{\beta \nu}}\}$ ($\{Y_i^{H_{\beta}}\}$) is the one that solves the equilibrium conditions
\begin{align}
    &\mu_C^H+\mu_e^H-\mu_{\nu_e}^H=0  \,\,\,\,\,\,\, (\mu_C^H+\mu_e^H=0)\\
    & \mu_S^H=0
\end{align}
together with the electromagnetic charge neutrality condition ($Y_C^H-Y_e^H=0$). By replacing the conditions in Equations~(\ref{eq:muCmue}) and (\ref{eq:muS}) and using Equation \eqref{eq:neutrinosglobal} (Equation \eqref{eq:neutrinosuntrapped}), we~obtain
\begin{align}
     &\mu_C^{Q*}+\mu_e^{Q*}-\mu_{\nu_e}^{H*}=\mu_C^{Q*}+\mu_e^{Q*}-\mu_{\nu_e}^{Q*}=0 \,\,\,\,\,\,\, (\mu_C^{Q*}+\mu_e^{Q*}=0)\\
     &\mu_S^{Q*}=0,
\end{align}
which, together with Equation \eqref{eq:YeQYCQcn}, are the conditions of $\beta$ equilibrium for quark matter (but with a different leptonic fraction with respect to the initial hadronic phase $Y_L^{Q*}\neq Y_L^H$ in the neutrino-trapped case). Thus, $Y_i^*=Y_i^{Q_{\beta}}$ is actually the choice for $\{Y_i^*\}$ that minimizes $W$ (and maximizes $\Gamma$) if the hadronic system is initially in neutrinoless $\beta$ equilibrium.

\subsection{{Summary}}
In conclusion, let us summarize our~assumptions:
\begin{itemize}
    \item The total nucleation probability is the product of (1) the probability $\mathcal{P}_2=\exp(-W_2/T)$ for a $Q_*$ droplet to be formed in an $H_*$ subsystem with the same composition $Y_i^{Q*}=Y_i^{H*}\equiv Y_i^*$ ($i=C,S$) and a baryon number $N_B^{Q*}=N_B^{H*}\equiv N_B^*$ and (2) the probability $\mathcal{P}_1=\exp(-W_1/T)$ that a subsystem $H_*$ exists in the system $H$.
    \item Since the volume of the deconfined quark droplet is typically much smaller than the total volume $V_{Q*}\ll V$, the~surroundings of the fluctuated hadronic subsystem $H_*$ and the quark droplet $Q_*$ act as a thermodynamical bath (Equations (\ref{eq:PHtPH}--\ref{eq:muiHtmuiH})).
    \item The composition of the quark droplet $Q_*$ is locally conserved from the hadronic fluctuated subsystem $H_*$ (Equations (\ref{eq:localYC}) and (\ref{eq:localYS})). 
    \item Since the nucleation probability is proportional to a negative exponential of $W/T$, we consider the fluctuated composition $\{Y_i^*\}$ that minimizes $W$ and thus will dominate the total probability (Equations (\ref{eq:muCmue}) and  (\ref{eq:muS})).
    \item We assume local electromagnetic charge neutrality in all of the phases (Equations (\ref{eq:YeQYCQcn}) and  (\ref{eq:YeHYCHcn})).
    \item We assume that in the fluctuation of the hadronic composition in the subsystem $H_*$, the~thermal and mechanical equilibrium with the surroundings are instantaneously restored (Equations (\ref{eq:THeq}) and (\ref{eq:PHeq})).
    \item Neutrinos are globally conserved and in equilibrium with the rest of the matter (if trapped) or neglected (if untrapped) (Equation (\ref{eq:neutrinosglobal}) or Equation (\ref{eq:neutrinosuntrapped})).
    \item The configuration with the critical deconfined droplet is a saddle point configuration (Equations (\ref{eq:tofindTQs}--\ref{eq:critRsaddle})). (This condition can be approximately replaced with the small degrees of metastability and the thermal equilibrium between the quark droplet and the surroundings.)
    \end{itemize}

\noindent Given 
 these assumptions, we found the following activation energy, namely the potential barrier height:
\begin{equation}
    W(n_B^H,\{Y_i^H\},T)=\frac{16 \pi}{3}\frac{\sigma^3}{\left[P_{Q}(n_B^{Q*},\{Y_i^{Q*}\},T)-P_H(n_B^H,\{Y_i^H\},T)\right]^2},
\end{equation}
where $n_B^{Q*},\{Y_i^{Q*}\}$ ( $i=C,S,e,\nu_e$) are fixed with
\begin{align}
    \mu_B^{Q}(n_B^{Q*},\{Y_i^{Q*}\},T)&= \mu_B^{H}(n_B^H,\{Y_i^{H}\},T) \label{eq:finalmuB}\\
    \mu_C^{Q}(n_B^{Q*},\{Y_i^{Q*}\},T)+\mu_e^{Q}(n_B^{Q*},Y_e^{Q*},T)&= \mu_C^{H}(n_B^H,\{Y_i^{H}\},T)+\mu_e^{H}(n_B^H,Y_e^{H},T)\label{eq:finalmuC}\\
    \mu_S^{Q}(n_B^{Q*},\{Y_i^{Q*}\},T)&=\mu_S^{H}(n_B^H,\{Y_i^{H}\},T) \label{eq:finalmuS}\\
      \mu_{\nu}^{Q}(n_B^{Q*},Y_{\nu_e}^{Q*},T)&= \mu_{\nu}^{H}(n_B^H,Y_{\nu_e}^{H},T) \label{eq:finalmunu}\\
    Y_C^{Q*}-Y_e^{Q*}&=0\label{eq:finalcn}.
\end{align}
where Equation \eqref{eq:finalmunu} reduces to $Y_{\nu}^{Q*}=Y_{\nu}^{H}=0$ if the neutrinos are not trapped.
Similarly, using the small degrees of metastability approach instead of the saddle point approach, the~activation energy is
\begin{equation}
    W(n_B^H,\{Y_i^H\},T)=\frac{16 \pi}{3}\frac{\sigma^3}{\left\{n_B^{Q*}\left[\mu_B^{Q}(n_B^{Q*},\{Y_i^{Q*}\},T)-\mu_B^H(n_B^H,\{Y_i^H\},T)\right]\right\}^2},
\end{equation}
where $n_B^{Q*},\{Y_i^{Q*}\}$ ($i=C,S,e$) are fixed with Equations~(\ref{eq:finalmuC}), (\ref{eq:finalmuS}), and (\ref{eq:finalcn}) and
\begin{equation}
    P_{Q*}(n_B^{Q*},\{Y_i^{Q*}\},T)= P_{H}(n_B^H,\{Y_i^H\},T) \label{eq:finalP}.\\
\end{equation}

\noindent The thermodynamical work $W_1$ needed to generate the fluctuated subsystem $H_*$ is
\begin{adjustwidth}{-\extralength}{0cm}
\begin{equation}
    W_1(n_B^H,\{Y_i^H\},T)=\frac{4}{3}\pi R_*^3 n_B^{Q*}\sum_{i=B,C,S,e}\left[Y_i^{Q*}\left(\mu_i^{H}(n_B^{H*},\{Y_i^{Q*}\},T)-\mu_i^{H}(n_B^{H},\{Y_i^H\},T)\right)\right],
\end{equation}
\end{adjustwidth}
where $R_*$ is
\begin{equation}
    R_*(n_B^H,\{Y_i^H\},T)=\frac{2\sigma}{ P_{Q}(n_B^{Q*},\{Y_i^{Q*}\},T)- P_{H}(n_B^H,\{Y_i^H\},T)} \label{eq:finalRsaddle}
\end{equation}
in the saddle-point approach and~\begin{equation}
    R_*(n_B^H,\{Y_i^H\},T)=\frac{2\sigma}{ n_B^{Q*}\left[\mu_B^{H}(n_B^H,\{Y_i^H\},T)-\mu_B^{Q}(n_B^{Q*},\{Y_i^{Q*}\},T)\right]}
\end{equation}
in the low-metastability approach, where  $n_B^{Q*},\{Y_i^{Q*}\}$ are computed as shown before and $n_B^{H*}$ is fixed using the condition {in Equation \eqref{eq:PHeq}}:
\begin{equation}
    P_{H}(n_B^{H*},\{Y_i^{Q*}\},T)=P_H(n_B^H,\{Y_i^H\},T).
\end{equation}

{We have thus computed the activation energy $W$ as a function of the initial hadronic conditions $n_B^H,\{Y_i^H\},T$, which could be, for example,~the thermodynamic conditions in the innermost region of a PNS or in a cell grid of a BNSM or CCSN simulation (which have a typical size much larger than the critical SQM droplet).}


\section{Color Superconductivity in~Nucleation}
\label{sec:fwork}

It has been suggested that the color superconductivity~\cite{Alford:2007xm} may play a fundamental role in compact stars and, in~particular, that quark matter, if~formed in compact stars, should be in a superconducting phase if the diquark pairing is large enough~\cite{Bombaci:2020vgw,Blaschke:2022egm,Ivanytskyi:2022qnw}.

A fundamental point to address is the behavior of color superconductivity in small systems, such as the first SQM droplet. Since the typical size (or coherence length) of a diquark pair is $\sim$$1/\Delta$ \cite{Alford:2007xm}, we expect a color-superconducting phase to form only in sufficiently large systems where diquark pairs have adequate space to develop (see also~\cite{Amore_2002}).
Thus, we will assume that the SQM will enter a {color-superconducting} phase only when the droplet of SQM has a radius $R$ larger than a specific threshold $R_{\Delta}$. 
{In this work, we will consider for simplicity the color flavor-locking (CFL) phase as the only color-superconducting phase of deconfined quark matter. Of~course, a~complete discussion should explore all possible color-superconducting phases~\cite{Alford:2007xm} (see, for example,~\cite{Bombaci:2006cs} for the nucleation of the 2SC phase).
However, in~this context, we are not interested in the details of the color-superconducting phase but~only  exploring the impact on quark nucleation in the case in which there is a more stable quark phase (the color-superconducting one), which can form only when the droplet is large enough (larger than the coherence length of diquark pairs). Below~a certain size,  the droplet is instead in a less stable quark phase (the unpaired one). This scheme can be applied to all possible color-superconducting phases. }

Within this framework, the~pressure (as well as the other thermodynamic quantities) becomes radius-dependent:
\begin{equation}
 P_Q(n_B^Q,\{Y_i^Q\},T,R) = 
 \begin{cases}
     P_{Qunp}(n_B^Q,\{Y_i^Q\},T)\,\,\,\,\,\,\,\,\text{ if }\,\,\,R \leq R_{\Delta}\\
        P_{QCFL}(n_B^Q,T)\,\,\,\,\,\,\,\,\text{ if }\,\,\,R > R_{\Delta}
 \end{cases} \label{eq:framPQ},
\end{equation}
where the details of the unpaired and CFL EOSs are reported in  Appendix \ref{sec:eosqua}.
In principle, one could define a smooth transition instead of such a step function. However, a~smooth approach would add complexity with no particular benefits, as~the results in the smooth approach could be easily achieved by adjusting the free parameters in the step approach. A~natural choice for $R_{\Delta}$ will be the coherence length. We will thus fix \mbox{$R_{\Delta}= \hbar c/\Delta(T)$}, where $\Delta(T)$ is the temperature-dependent diquark gap (see \mbox{Appendix \ref{sec:eosquacfl}}).

This approach can be applied to situations where the energetic convenience of the bulk, compared with a finite system, arises even from factors other than surface tension. In~terms of the Bag model, we can describe this possibility by assuming that the Bag parameter depends on the volume of the SQM system (or, equivalently, the~baryon number). In~particular, the~Bag would decrease as a function of the baryon number of the~system.

The critical droplet radius (Equation (\ref{eq:finalRsaddle})) becomes
\begin{adjustwidth}{-\extralength}{0cm}
\begin{equation}
    R_*(n_B^H,\{Y_i^H\},T)=
    \begin{cases}
        R_{unp*}(n_B^H,\{Y_i^H\},T)\,\,\,\,\,\,\,\,\,\,\,\,\,\,\,\,\,\,\,\,\,\,\,\,\,\,\,\,\,\,\,\,\,\,\, \text{ if }\,\,\, R_{unp*}(n_B^H,\{Y_i^H\},T) \leq R_{\Delta}\\
        \max\left[R_{\Delta},  R_{CFL*}(n_B^H,T) \right] \,\,\,\,\,\,\,\,\,\,\,\,\,\,\,\,\,\,\,\,\,\,\,\, \text{ if }\,\,\, R_{unp*}(n_B^H,\{Y_i^H\},T) > R_{\Delta}
    \end{cases}
   \label{eq:framRs}.
\end{equation}
\end{adjustwidth}

\section{Results and~Discussion}
\label{sec:results}

In this section, we will report and discuss the results of our analysis. 
The primary focus will be on discussing the qualitative and quantitative effects of thermal fluctuations on the matter composition and formation of color-superconducting gaps during the nucleation process in the three-flavor case.
In particular, the~potential barrier between the metastable hadronic phase and the stable SQM phase will be investigated using different~approaches:
\begin{itemize}
    \item Unpaired+
CFL: The complete framework is described in Section~\ref{sec:fwork}. 
    The CFL phase fulfills the Witten hypothesis, but it
    can be formed only if the SQM droplet is larger than the typical diquark coherence length $R_{*}\geq R_{\Delta}=1/\Delta(T)$. The~composition $\{Y_i^*\}$ fluctuates with respect to the average bulk one $\{Y_i^H\}$, and it is fixed to maximize the nucleation rate $\Gamma$. (All other values of  $\{Y_i^*\}$ are subleading for the calculation of $\Gamma$ and can be neglected.) Thus, the~$Q_*$ thermodynamical quantities are fixed using Equations~(\ref{eq:finalmuB}--\ref{eq:finalcn}) for the unpaired phase and Equations~(\ref{eq:tofindnBQs}) and (\ref{eq:finalmunu}) for the CFL phase.
    \item Unpaired: This is as described before but~without the CFL phase. It corresponds to the case where  $R_{\Delta}\rightarrow +\infty$. 
    \item Frozen composition: The thermal fluctuations of the hadronic composition are neglected. The~$Q_*$ thermodynamical variables are computed by fixing $Y_i^*=Y_i^H$ ($i=C,S,e$ or, equivalently, $i=u,d,s,e$) and Equations~(\ref{eq:tofindnBQs}) and (\ref{eq:finalmunu}). This is the approach used in~\cite{Bombaci:2016xuj}. 
\end{itemize}

\noindent In the examples reported in this section, we will use a parametrization for the CFL phase that fulfills the Witten hypothesis on the absolute stability of SQM~\cite{Witten:1984rs,Bodmer:1971we} and a parametrization for the unpaired phase that does not lead to absolutely stable SQM, namely $E/A_{CFL}<E/A_{^{56}Fe}<E/A_{unp}$ in bulk at zero pressure and temperature (see \mbox{Appendix \ref{sec:eosqua}}). This choice allows for the existence of compact stars as massive as $\sim$$2.6$~M$_{\odot}$ and which are interpreted as being CFL-QSs in the context of the two-family compact stars scenario~\cite{Bombaci:2020vgw}. At~the same time,  for~the unpaired phase, the~potential barrier for the nucleation turns out to be large enough to shift the threshold of formation of this phase at large values of $n_B^H,T$ to make the appearance of SQM at an $n_B^H,T$ value which is not too low. A~detailed discussion will be provided in a future work (in preparation).

However, we stress that the framework presented here can also be applied in cases in which both the CFL and unpaired SQM do not fulfill Witten's~hypothesis. 

\begin{figure}[H]
\begin{adjustwidth}{-\extralength}{0cm}
\centering
\includegraphics[width=1.\textwidth]{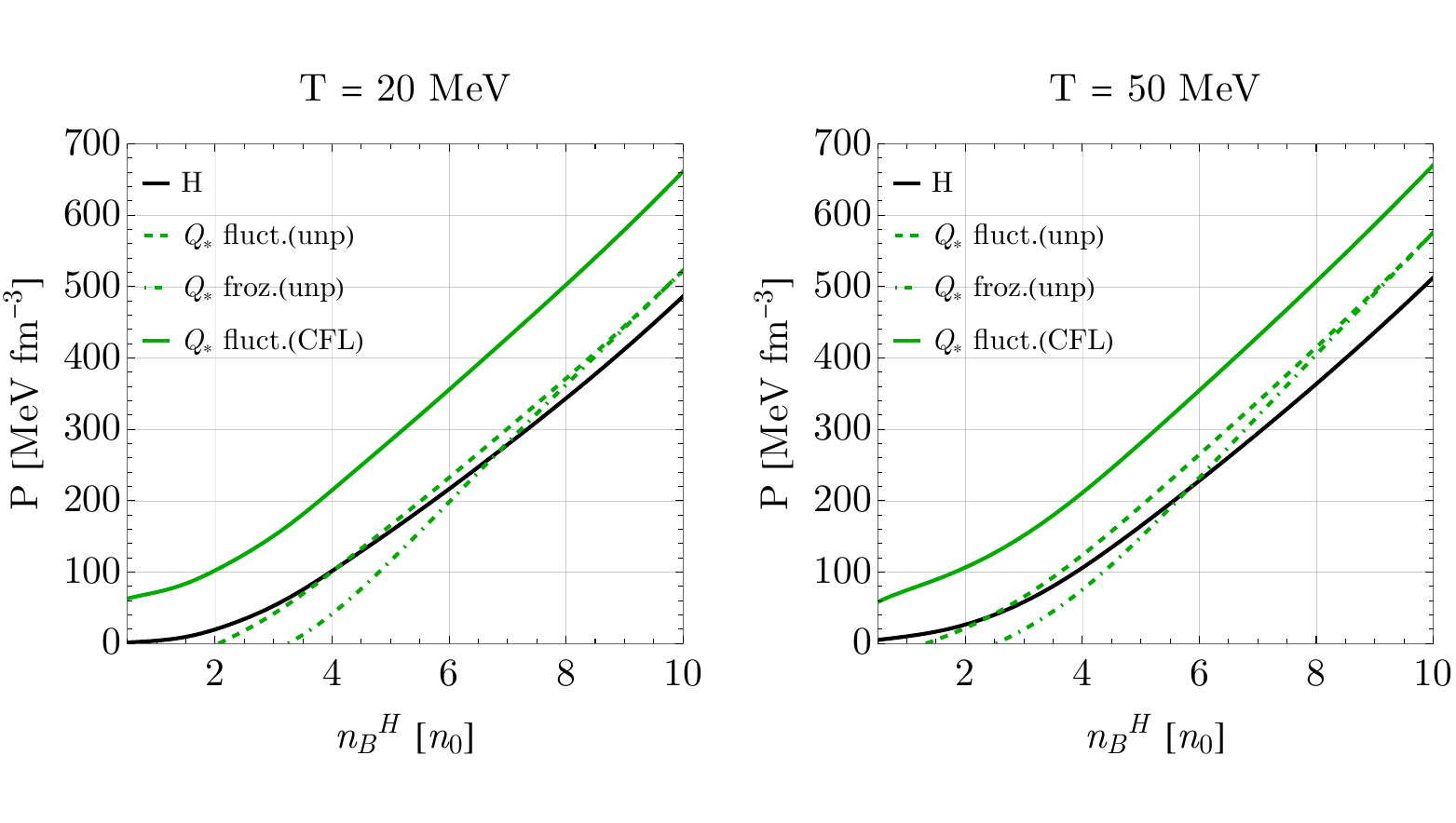}\vspace{6pt}
\includegraphics[width=1.\textwidth]{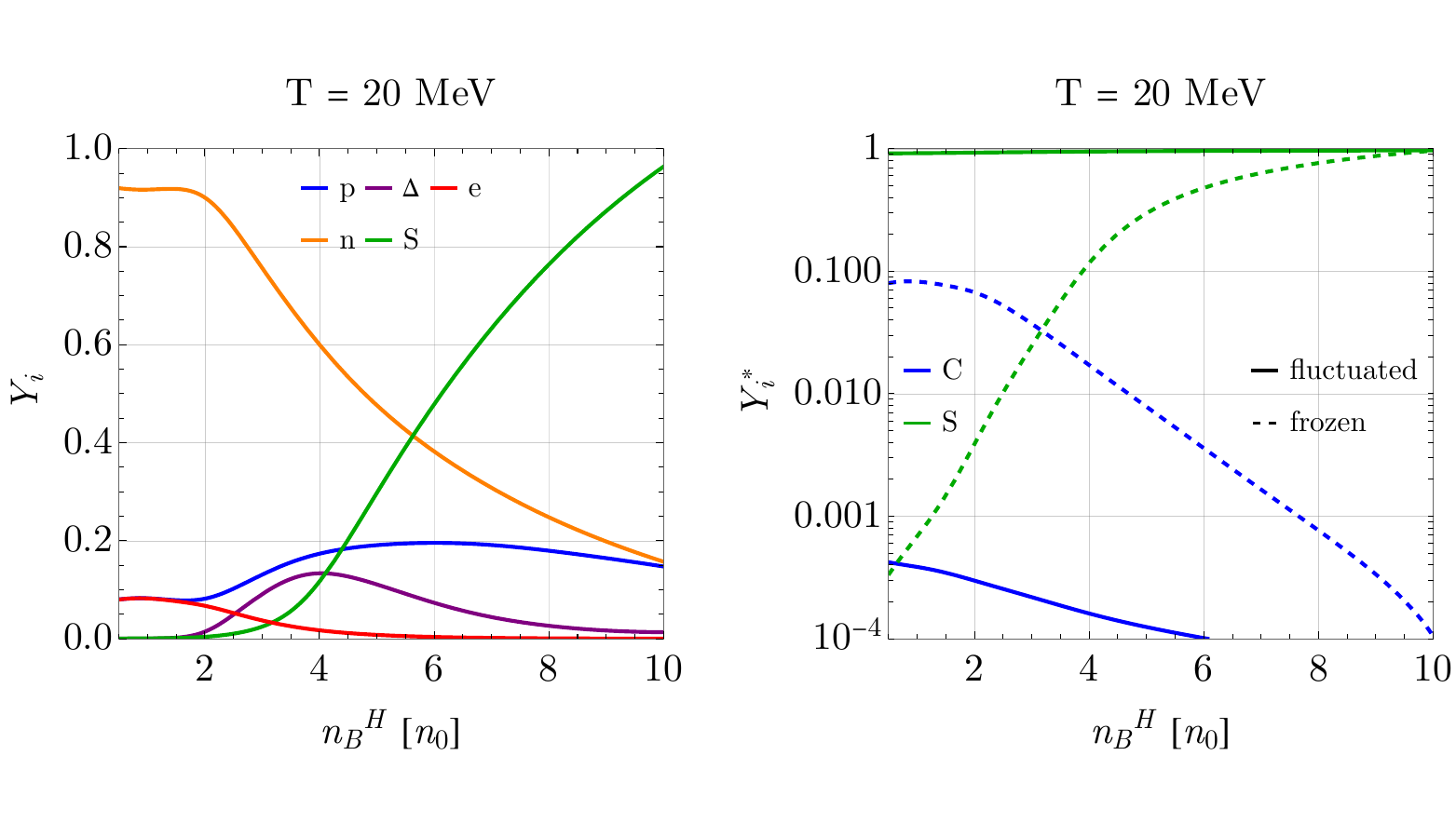}
\end{adjustwidth}
\caption{(\textbf{top}) Pressure $P$ as a function of the initial baryon density $n_B^H$  at a fixed temperature $T=20$~MeV (\textbf{left}) and $T=50$ MeV (\textbf{right}). The~different lines refer to the hadronic phase $H$, unpaired SQM with a fluctuated composition ($Q_*$ fluct. unp), unpaired SQM with a frozen composition  ($Q_*$ froz. unp), and CFL SQM ($Q_*$ fluct. CFL). 
(\textbf{bottom}) Composition of the initial hadronic phase $\{Y_i^H\}$ as a function of the initial baryon density $n_B^H$  at a fixed temperature $T=20$ MeV (\textbf{left}). For~easier interpretation of the plot, hyperons are not shown 
 separately, but~the total strangeness is reported. The~contribution of the deltas $\Delta^{++},\Delta^{+},\Delta^{0},\Delta^{-}$ are summed for $\Delta$. Composition of the first droplet of unpaired SQM in the fluctuated and frozen cases (\textbf{right}). The~frozen composition corresponds to the initial hadronic one.
The initial hadronic phase is in $\nu$-less $\beta$ equilibrium. We fixed the parameters to $B_{unp}^{1/4}=175$ MeV, $B_{CFL}^{1/4}=135$ MeV , and $\Delta_0=80$ MeV.
\label{fig:EOS}}
\end{figure}

Some properties of the used EOSs are reported in Figure~\ref{fig:EOS}. In~particular, the~upper panels show the pressure as a function of the initial hadronic baryon density $n_B^H$ at two different temperatures for the initial hadronic phase ($H$), for~the unpaired SQM with the fluctuated composition ($Q_*$ fluct. unp), for~the unpaired SQM computed while assuming a frozen composition ($Q_*$ froz. unp), and for the CFL SQM ($Q_*$ fluct. CFL).   
Note that the three $Q_*$ pressures are not reported as a function of their baryon densities $n_B^{Q*}$ since they were computed by imposing Equations~(\ref{eq:finalmuB}--\ref{eq:finalcn}) at given $n_B^H$ and $T$ values. 
The favored phase is the one with the highest pressure at given $n_B^H$ and $T$ values (see Equation~(\ref{eq:critWsaddle}), where the bulk part of $W$ is the energy difference in bulk between  phases $Q_*$ and $H$.
In our scenario, the~CFL phase fulfilled the Witten hypothesis, and~it was indeed always more stable in bulk than the hadronic phase. On~the other hand, unpaired SQM did not fulfill the Witten hypothesis in our example and became more stable than the hadronic phase only at high enough $n_B^H$ or $T$ values. The~intersection points $P_{Q*}=P_{H}$ correspond to the coexistence pressure of the mixed phase in bulk. If~$\sigma=0$, then no finite-size effects take place, and the deconfinement starts as soon as the bulk coexistence pressure is reached. However, if~$\sigma \neq 0$, then the~critical radius $R_*^{unp}$ diverges at $P_H=P_{Q*}$, as can be noted in Equation \eqref{eq:finalRsaddle}. At~the $n_B^H$ and $T$ values at which $P_{Q*}>P_{H}$, the~hadronic phase is metastable with respect to the $Q_*$ phase, $R_*^{unp}$ has a finite value, and~the unpaired SQM nucleation has a finite probability to happen.
The $Q_*$ phase with the composition fluctuation that minimizes $W$ is always energetically favorable with respect to the one with a frozen composition. At~the fixed initial hadronic phase, with~values $n_B^H,\{Y_i^H\},T$, the~total work $W$ (Equation (\ref{eq:critWsaddle})) depends only on $Q_*$. Since the bulk term of $W$ is the energy difference between the $Q_*$ and $H$ phases in bulk, the~most stable $Q_*$ in bulk is the one that minimizes $W$.

In the results presented here, the~initial hadronic phase was in $\nu$-less $\beta$ equilibrium ($H_{\beta}$). Note that even if the initial hadronic phase were in equilibrium with respect to weak reactions, weak interactions do not play any role during  nucleation. Moreover, we stress that our framework can be easily applied to an initial hadronic phase with a generic composition $\{Y_i^H\}$ ($i=C,S$) while not assuming any weak equilibrium. The~model for computing the hadronic EOS is reported in Appendix \ref{sec:eoshad}.

By increasing the temperature, the~unpaired SQM pressure increased faster than the hadronic pressure, both in the fluctuated and frozen cases. This implies that the coexistence pressure is reached at a lower value of $n_B^H$. The~behavior of the CFL pressure was not monotonic with the temperature since $\Delta$ decreased with the~temperature. 

The bottom panels show the hadronic composition (left) and the unpaired\endnote{The CFL composition is not shown 
 since it is always $Y_C^{*}=0$ and $Y_S^*=1$.} $Q_*$ composition in the fluctuated and frozen cases (right) as a function of the initial hadronic baryon density and at fixed temperature.
A discussion about the roles of hyperons and deltas can be found in~\cite{Drago:2013fsa,Drago:2014oja}.
The fluctuated case, namely the composition $\{Y_i^*\}$ of the fluctuated subsystem associated with the lowest potential barrier height, showed extremely high values for $Y_S^*$ and low values for $Y_C^*$ compared with the values in the hadronic phase in~the low-$n_B^H$ regime. By~increasing $n_B^H$, hyperons and deltas appeared, $Y_S^H$ ($Y_C^H$) increased (decreased), and~the frozen composition became more and more similar to the fluctuated one.
As noted in Section~\ref{sec:nucleation}, as~long as the initial hadronic phase is in $\beta$ equilibrium, the~unpaired fluctuated composition corresponds to the composition of $\beta$ equilibrium SQM. This consideration applies whether neutrinos are trapped or not; however, in~the former case, $Y_L^{Q*}\neq Y_L^H$. 
We stress again that the compositions are always reported as a function of $n_B^H$ and not  the baryon density of the $H_*$ or  $Q_*$ phases.

\begin{figure}[H]
\begin{adjustwidth}{-\extralength}{0cm}
\centering
\includegraphics[width=1.1\textwidth]{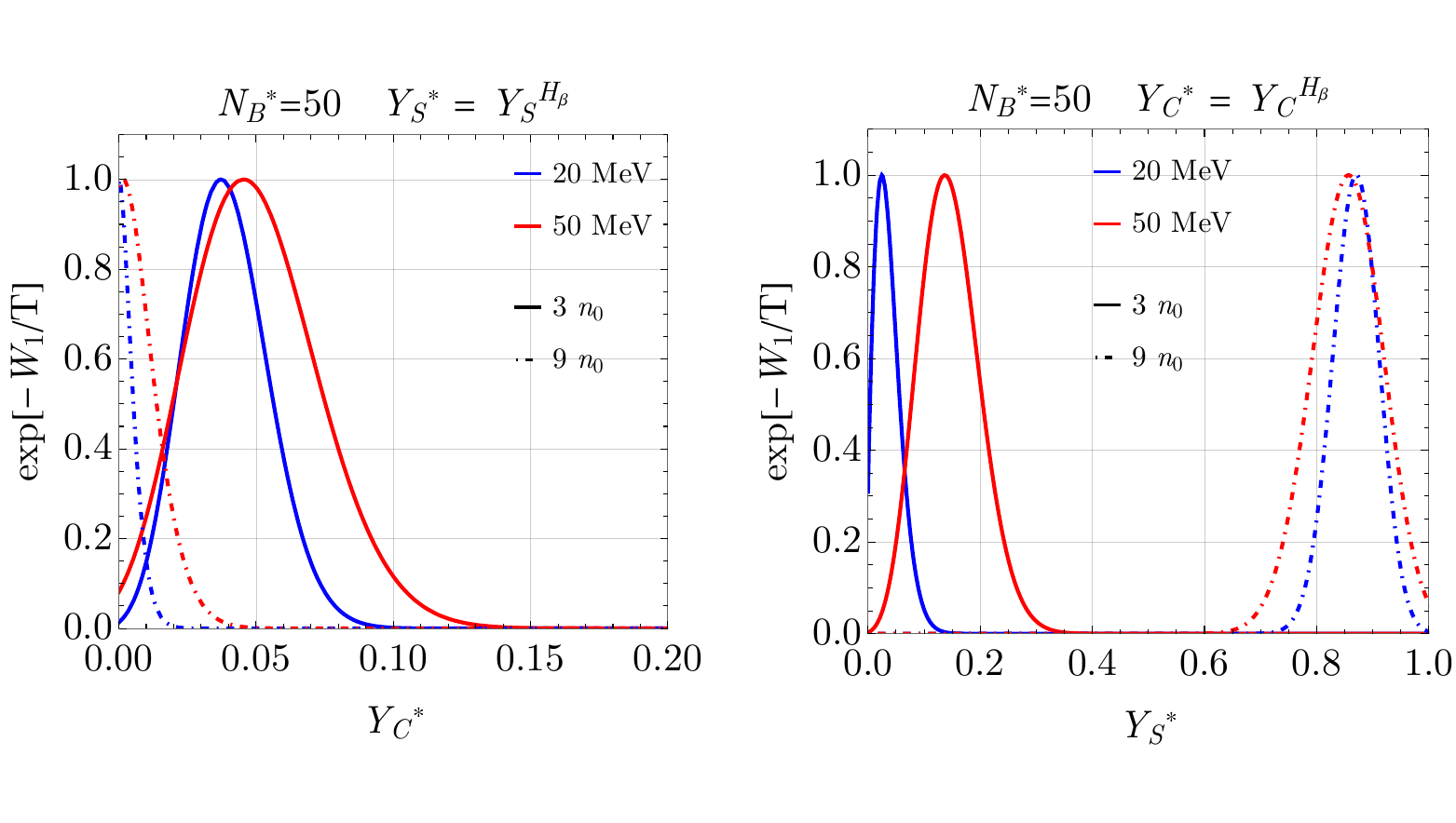}
\end{adjustwidth}
\caption{Non-normalized (see~\cite{Guerrini:2024gzu} for a discussion on the normalization) probability for a subsystem characterized by a baryon number $N_B^*$ and in thermal and mechanical equilibrium with the surroundings to have a certain $Y_C^*$ value at a fixed $Y_S^*=Y_S^{H_{\beta}}$ (\textbf{left}) and a certain $Y_S^*$ value at a fixed $Y_C^*=Y_C^{H_{\beta}}$ (\textbf{right})\label{fig:gauss}.
Two different temperatures $T$ and baryon densities $n_B^H$ are shown.}
\end{figure}

Figure~\ref{fig:gauss} displays the composition fluctuation probability ($\exp(-W_1/T)$). A~complete discussion on the normalization and a comparison with a multivariate Gaussian can be found in~\cite{Guerrini:2024gzu}. In~the left (right) panel, the composition fluctuation probability is shown as a function of $Y_C^*$ ($Y_S^*$) with a fixed $Y_S^*=Y_S^{H_{\beta}}$ ($Y_C^*=Y_C^{H_{\beta}}$). The~maximum of the distributions corresponded to $W_1=0$ MeV, namely  the average equilibrium configuration $Y_i^*=Y_i^{H_{\beta}}$ ($i=C,S$). At~$T=0$, the~thermal fluctuations were absent, and the distribution was a $\delta$ function centered at the equilibrium configuration. However, at~a finite temperature, a~subsystem characterized by $N_B^*$ baryons could locally have a composition different from the average equilibrium one. The~higher the temperature, the~higher the probability of big fluctuations in the composition (and namely, the~wider the distribution around the peak). Moreover, by~increasing the size of the subsystem in terms of $N_B^*$, the~distributions shrank around the maximum. This behavior is straightforward; when fixing the fluctuation in terms of the composition $\Delta Y_i$, the~larger $N_B^*$ is, the~larger the fluctuation in terms of the number of particles $\Delta N_i$. 
At $n_B^H=3\, n_0$, and $T=20$ MeV, when assuming the initial hadronic phase to be in $\nu$-less $\beta$ equilibrium, the~average strangeness fraction was $Y_S^H \simeq 0.02$. In~these conditions, the~probability of having a subsystem with the composition that maximizes the rate $Y_S^*\simeq 0.9$ is extremely low ($ \exp(-W_1/T)\sim 10^{-94}$). More discussion on $N_B^*$ and on the probabilities will be carried out after  this~section.

\begin{figure}[H]
\includegraphics[width=0.8\textwidth]{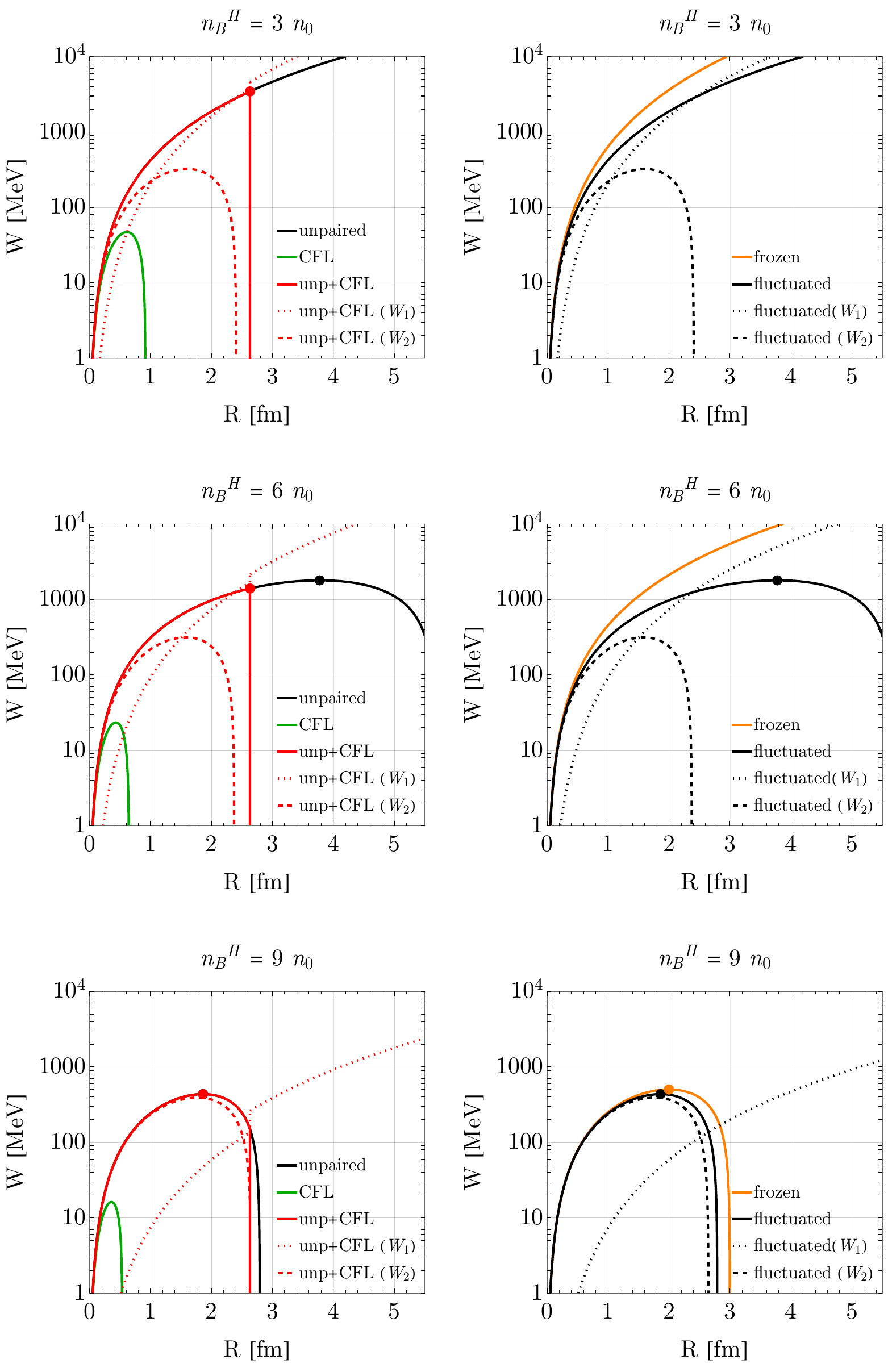}
\caption{{Thermodynamical work $W$ needed to generate an SQM droplet as a function of the droplet radius $R$ (namely, the~energy barrier separating the metastable hadronic phase and the stable SQM phase) for $n_B^H=3\,n_0$ (\textbf{top}), $n_B^H=6\,n_0$ (\textbf{middle}), and  $n_B^H=9\,n_0$ (\textbf{bottom}). 
In the left column, the red line refers to the framework in which SQM is in the CFL (unpaired) phase if the droplet is larger (smaller) than the coherence length of the diquark pairs (unp+CFL; see Section~\ref{sec:fwork}), the~black line considers only unpaired SQM, and the green line considers only CFL SQM. In~all three cases, thermal fluctuations in the composition were considered.
The dashed and dotted lines refer to $W_2$ (work needed to convert a subsystem $H_*$ to $Q_*$ with $H$ surroundings) and $W_1$ (work needed to generate a subsystem $H_*$ in a system $H$), respectively. The~red dot denotes the saddle point configuration, namely the critical radius $R_*$ and the activation energy $W(R_*)$, while the black dot refers to the critical radius in the unpaired-only case $R_*^{unp}$.
In the right column, the orange lines refer to the frozen composition case (specifically neglecting thermal fluctuations in the composition), while the black lines are the same as those in the left column. The~solid, dotted, and~dashed black lines show $W$, $W_1$, and~$W_2$ in the fluctuated unpaired case, respectively.
The initial hadronic matter was in $\nu$-less $\beta$ equilibrium. We set $T=20$ MeV, $\sigma=30$ MeVfm$^{-2}$, $\Delta_0 = 80$ MeV, and $B_{unp}^{1/4}=175$ MeV.}\label{fig:WR} } 
\end{figure}

Figure~\ref{fig:WR} shows the work needed to generate an SQM droplet as a function of the droplet radius $R$ at a fixed temperature $T$ and initial hadronic baryon density $n_B^H$.
The maximum for the work was reached when the system was fully in the hadronic phase, except~for a critical droplet with a critical radius $R_*$ (such that $\max[W(R)]=W(R_*)$). Such a configuration corresponds to a saddle point, characterized by a maximum in the $W-R$ plane (unstable mechanical equilibrium point) and minima for the other independent variables (see \mbox{Section~\ref{sec:nucleation}}).
The value $W(R_*)$ can be interpreted as the potential barrier height separating the metastable hadronic phase from the stable SQM phase. 
In the left column's panels, the~different lines were obtained by computing the work using Equation \eqref{eq:critWsaddle} but with different EOSs for the SQM phase; the green line was obtained using $P_{Q*}=P_{QCFL}(n_B^{Q*}, T)$ (CFL-only), the~black lines were obtained using $P_{Q*}=P_{Qunp}(n_B^{Q*},\{Y_i^*\}, T)$ (unpaired-only), and the red lines were obtained using the framework described in Equation \eqref{eq:framPQ} (unp+
CFL). In~this framework, for~$R<R_{\Delta}$, the~SQM was in an unpaired phase (black and red lines are identical) since diquark pairs had not formed, as~their typical coherence length ($\sim 1/\Delta$) was larger than the droplet. When $R>R_{\Delta}$, diquark pairs formed, leading to the superconducting phase (red and green lines are identical).
The critical radius in the CFL-only case $R_{*}^{CFL}$ was always smaller than the unpaired-only one $R_{*}^{unp}$. We thus obtained two qualitatively different scenarios. If the critical radius in the unpaired-only case (the maximum point of the black curve) was smaller than $R_{\Delta}$, then the~critical radius in the unp+CFL case corresponded to the unpaired-only one, and~the saddle point configuration was reached before the diquark pairs formed ($n_B^H=9 \, n_0$); otherwise, if~the maximum in the unpaired-only case was not reached before $R_{\Delta}$ ($n_B^H=6\,n_0$) or not reached at all ($n_B^H=3\,n_0$), then diquark pairs started to form as soon as $R\geq R_{\Delta}$, the~unp+CFL work dropped to the CFL-only work, and~the maximum for the work was thus at $R_*=R_{\Delta}$ (left panel). 
By choosing a larger value for $\Delta_0$,~$R_{*}^{unp}$ was not affected, while $R_{\Delta}$ became lower, leading to a smaller $R_*$ when $R_{*}^{unp}>R_{\Delta}$.

The dashed and dotted red lines show $W_2$, namely the work needed to generate a droplet $Q_*$ in a subsystem $H_*$, and~$W_1$, namely the work needed to generate a hadronic subsystem $H_*$ characterized by a fluctuated composition with respect to the average bulk value $H$, respectively. The~work $W_1$ is always positive since the energy of a fluctuated hadronic configuration is always larger than the equilibrium configuration. Moreover, it monotonically increases with $R$ since the larger the fluctuation (namely, the~larger \mbox{$N_B^{*}=n_B^{Q*}V_{Q*}\propto R^3$}), the~larger the energy difference between the fluctuated and  equilibrium configurations.
On the other hand, $W_2$ contains two terms; the finite size term is always positive and monotonically increasing with $\propto R^2$, while the bulk term is negative (positive) and monotonically decreasing (increasing) with $\propto R^3$ if   $Q_*$ is more (less) stable than $H_*$. Thus, if~$Q_*$ is more stable than $H_*$, then $W_2$ is positive at low $R$ values, reaches a maximum in $R_2$, and~becomes negative for large $R$ values. The~discontinuities at $R=R_{\Delta}=1/\Delta(T)$ are due to the onset of the CFL phase (which is sharp in our simplified approach).
The work $W_1$ is smaller at $n_B^H=9\, n_0$ since the difference between the composition $\{Y_i^H\}$ and $\{Y_i^{Q*}\}$ is much smaller than the one at $n_B^H=3\, n_0$, as can be easily noted in Figures~\ref{fig:EOS} and \ref{fig:gauss}.

When increasing $n_B^H$ and $T$ at fixed $R$ values, $W \propto (P_H-P_{Q*})$ decreased (see top panels in \mbox{Figure~\ref{fig:EOS}}), leading to a lower activation energy $W(R_*)$ even if $R_*$ did not change (i.e., even if \mbox{$R_{unp}^*>R_{\Delta}$}).
Note that under the conditions in the top left panel ($n_B^H=3\,n_0$, and $T=20$ MeV), $H$ was not metastable with respect to the unpaired $Q_*$ phase, and thus the black line monotonically increased and $R_*^{unp}$ was not defined. However, in~the CFL+unp framework, nucleation is, in principle, still possible thanks to the onset of the CFL phase at $R=R_{\Delta}\simeq 2.6$ fm. 

The panels in the right column show the work as a function of the droplet radius for the unpaired SQM with a fluctuated (black solid line) and frozen (orange solid line) composition. Note that the black solid lines in the two columns are identical.
In general, we note that the potential barrier $W(R_*^{unp})$ was always lower in the fluctuated case. 
At $T=20$~MeV and $n_B^H=3\, n_0$, the~hadronic phase in $\nu$-less $\beta$ equilibrium was more stable than the unpaired $Q_*$ phase, both in the fluctuated and frozen cases (see the top left panel in Figure~\ref{fig:EOS}). Thus, both the solid orange and black lines  monotonically increased, and the critical radii did not exist. At~$n_B^H=6 \, n_0$, $H_{\beta}$ was less stable than $Q_*$ in the fluctuated case, which was thus not monotonically increasing and had a critical radius (black dot). Finally, $Q_*$ was more stable than $H_{\beta}$ in both the fluctuated and frozen cases at $n_B=9\,n_0$. Moreover, since in these conditions, $\{Y_i^{H_{\beta}}\}$ was similar to the $\beta$-equilibrium unpaired SQM, then the frozen and fluctuated cases were similar.
The dotted and dashed lines refer to $W_1$ and $W_2$ as shown in the left column but~considered only unpaired SQM. Note that $W_1=0$ and $W_2=W$ in the frozen composition~approach. 

In general, the~higher the surface tension $\sigma$, the~higher the potential barrier. In~particular, the~work $W$ at a fixed $R$ value depends linearly on the surface tension $W\propto \sigma$ (see \mbox{Equation~(\ref{eq:critWsaddle}})). The~dependence of $R_*$ and $W(R_*)$ on $\sigma$ was different in the two cases $R_{*}^{unp}<R_{\Delta}$ and $R_{*}^{unp}>R_{\Delta}$. In~the former case, $R_*=R_*^{unp}\propto \sigma$ (see Equation~(\ref{eq:critRsaddle})), and thus $W(R_*)\propto \sigma^3$ (see Equation~(\ref{eq:critWcsaddle})). On~the other hand, in~the latter case, $R_*=R_{\Delta}\not{\propto} \sigma$, and thus $W(R_*)\propto \sigma$.

Finally, we provide a qualitative discussion on the order of magnitude of the potential barrier height for the conditions relevant in astrophysical systems. In~particular, using \mbox{Equation \eqref{eq:nucltime}}, we note that the nucleation time is $\tau\simeq 1$ s when \mbox{$W/T \simeq (165$--
$180)$} for the astrophysically relevant range of $n_B^H$ and $T$. This large value leads to an extremely low nucleation probability $\exp(-W/T)\simeq 10^{-79}-10^{-72}$ and an extremely low nucleation rate \mbox{$\Gamma\simeq 2\times 10^{-52}$ s$^{-1}$fm$^{-3}$}. However, such a small rate is still enough to have a nucleation event typically in one second, since the system we are considering, namely a sphere with a radius of $100$~m in the compact star core, is rather big ($V\simeq 4\times 10^{51}$ fm$^3$). In~other words, even if the nucleation rate is locally quite small, only one event is needed to trigger deconfinement, and~in the system, there are many possible subsystems in which it can happen.
Moreover, we note that the values of $\exp[W_1/T]$ and $\exp[W_2/T]$ leading to a nucleation time of $\sim$$1$ s are quite different when changing $n_B^H,T$, while their product always leads to values of the order of magnitude reported above.  
For example, in~the unp+CFL fluctuated approach, at $n_B^H=3 \, n_0$ and $\sigma=30$ MeV fm$^{-2}$, $\tau=1$ s is obtained at $T\simeq20$ MeV, leading to $W/T\simeq172$, and~$\exp[-W/T]\simeq10^{-75}$. In~these conditions, we obtain $W_1/T\simeq 185$ and $W_2/T\simeq-12$. The~number of baryons for the critical droplet is $N_B^*\simeq45$. At~$n_B^H=6 \, n_0$, $\tau=1$ s is reached at $T\simeq8$ MeV, $W/T\simeq177$, $N_B^*\simeq68$, $W_1/T\simeq 254$, and $W_2/T\simeq -78$.
Thus, even if a certain subsystem characterized by a fluctuated composition has an extremely tiny probability to form (small $\exp[W_1/T]$), it could appear somewhere anyway in such a big system, and its nucleation probability could be greatly enhanced by the contribution of $\exp[W_2/T]$. Thus, the~total nucleation probability turns out to be much greater than the one obtained within the frozen composition approach.
At fixed values of $n_B^H=3\, n_0$, and $\tau=1$ s, the~temperature in the unpaired-only fluctuated and frozen cases was higher than that in the unp+CFL approach ($T\simeq50$~MeV in the unpaired-only fluctuated approach and $T\simeq80$ MeV in the unpaired frozen approach). The~activation energy divided by the temperature was similar in all cases ($W/T\simeq170$ and $W/T\simeq168$, respectively), even if $W_2/T$ and $W_1/T$ were quite different for different cases and thermodynamical conditions. When increasing $n_B^H$, the above-mentioned quantities converged to the same values for the three different approaches, since the unpaired and unp+CFL approaches became identical when $R_*^{unp}<R_{\Delta}$ and the frozen approach converged to the fluctuated one when the initial average hadronic composition matched the composition of $\beta$-equilibrium SQM.
{A more quantitative discussion as well as a detailed investigation on the impact of free parameters will be provided in a future work.}

\section{Summary and~Conclusions}
\label{sec:conclusions}
The main goal of this work was to present a framework that aims to investigate the role of thermal fluctuations in the composition and  color superconductivity in the nucleation of SQM in the typical conditions of high-energy astrophysical systems related to compact stars. 
In the context of deconfinement first-order phase transitions, nucleation can be interpreted as overcoming  the potential barrier between the metastable hadronic phase and the stable SQM phase, and~it takes into account the finite-size effects and the timescale of the fluctuations. 
Within the Witten hypothesis on the absolute stability of SQM and  the two-family scenario of compact stars, SQM nucleation is the trigger leading to the conversion of an ordinary NS into a QS. 
{However, it is important to stress that the formalism  presented here only applies to the nucleation and~cannot describe the subsequent evolution leading to the conversion into a QS, which can be qualitatively divided into different steps (see, for example,~\cite{Drago:2015dea}). (1) The critical droplet keeps expanding or merges with others until its size becomes macroscopic, and (2) further expansion of the macroscopic droplet of quark matter inside the hadronic star can be described by using hydrodynamical equations, and~it can be divided into  a rapid burning of the inner part first and a slow burning of the outer part~\cite{Herzog:2011sn,Drago:2015fpa}. }

While a complete treatment of the nucleation process would require microphysical calculations that consider interaction rates and diffusion timescales, an~approximate analysis can be performed within a thermodynamic approach. 
A key point is that the timescale associated with the weak interactions is much longer than that of strong interactions, which are responsible for the deconfinement phase transition.
Thus, a~standard approach is to impose that the newly formed quark matter droplets have the same composition as the initial hadronic phase, since weak interactions do not have sufficient time to modify the flavor composition during~nucleation. 

However, this approach neglects thermal fluctuations of the composition, assuming that it is locally identical everywhere in the system. 
Here, we presented explicit thermodynamic calculations in which the total nucleation probability is computed as the product of the nucleation probability in a subsystem $H_*$ characterized by a composition locally different from the average bulk of a system $H$ and~the probability that the subsystem $H_*$ exists in the system. This approach was already  described in our previous work~\cite{Guerrini:2024gzu}, and it has been reviewed in some parts and applied to the three-flavor case~here. 

Within the assumptions  described  here (and summarized at the end of Section~\ref{sec:nucleation}), we found that the first droplet of SQM had the same chemical potentials $\mu_C+\mu_e$ and $\mu_S$ of the initial hadronic phase. This also implies that if the initial hadronic phase is in $\beta$ equilibrium, then the first droplet of SQM will also be in $\beta$ equilibrium, possibly with a different $Y_L^{Q*}\neq Y_L^H$ in~case the neutrinos are trapped. We stress that this result would have been obtained even if weak interactions were not playing any role during nucleation. Thus, only finite size effects contributed to the potential barrier, while the chemical unbalance related to strong versus weak timescales was hidden due to thermal fluctuations in the~composition. 

We found that for~the typical temperatures reached in the CCSNe and PNS evolution, thermal fluctuations in the composition, in~general, lowered the potential barrier, leading the SQM to appear at lower $n_B^H$ and $T$ values. The~role of thermal fluctuations became negligible when the composition of the initial hadronic phase was similar to the $\beta$-equilibrium SQM~phase.

This result is also important for the standard one-family or twin-star scenario of compact stars, since the frozen composition approach {usually} led to a big delay in~terms of $n_B^H$ and $T$ for~triggering the formation of the mixed phase {(in particular in the
three-flavor case)}. On~the other hand, considering the thermal fluctuations of the composition, the~delay was downsized and related only to finite size effects (i.e., the~surface tension).

It has been suggested that color superconductivity could be crucial for explaining the phenomenology of compact stars and~the massive compact star observations in particular, both in hybrid stars in the standard one-family scenario (see, for example,~\cite{Gartlein:2023vif,Blaschke:2022egm,Ivanytskyi:2022qnw}) and in QS within the two-family scenario (see, for example,~\cite{Bombaci:2020vgw}). 
Since the typical coherence length of a diquark pair is $\sim$$1 /\Delta$, we propose a framework in which the SQM EOS can be described as unpaired if the SQM droplet is smaller than $1/\Delta$ and as a color-superconducting phase (we considered a CFL phase for simplicity) as soon as the quark droplet is larger than the coherence length of the diquark pairs.
Within this framework, the~maximum QS mass (or hybrid star mass) is set by the chosen parametrization of the CFL EOS, while the nucleation conditions depend on the unpaired EOS, the diquark coherence length, and~thus  the temperature-dependent superconducting gap as well $R_{\Delta}(T)=1/\Delta(T)$. 

The qualitative results within this framework depended on whether or not the critical radius of the unpaired SQM droplet was reached before that of the diquark pairs appeared (namely, if~$R_*^{unp}<R_{\Delta}$). In~particular, at~high $n_B^H$ or $T$ values, if $R_*^{unp}<R_{\Delta}$, then $R_*=R_*^{unp}$, and the results with unpaired SQM were reobtained . On~the other hand, at~low $n_B^H$ and $T$ values,  $R_*^{unp}>R_{\Delta}$, and the maximum of $W$ was reached as soon as diquark pairs appeared; thus, $R_*=R_{\Delta}$. Note that in the latter case, the~critical radius was independent of $n_B^H$ but depended on $T$ since $\Delta=\Delta(T)$.

We stress again that even if we  chose for the examples a CFL phase that respected the Witten hypothesis, the~same framework could be easily applied to hybrid stars in the one-family scenario, and~a similar qualitative discussion could be~carried out.

Our calculation employed a thermodynamic approach to model the decay of a metastable state into a stable state. However, this approach could misestimate the true nucleation probabilities. 
For example, the nucleation time could be underestimated since some microphysical mechanisms could make the composition fluctuation less likely by slowing strangeness~diffusion.

We assumed here a local electromagnetic charge neutrality, while a complete discussion should properly take into account how leptons distribute in space to screen the electric charge. {However, we plan to relax this assumption in a future work}. 

{Here, we adopted ``two-phase approximation'', assuming a first-order phase transition between the confined (hadronic) and deconfined (SQM) phases, each described by distinct equations of state. Since the potential for the order parameter of the phase transition that consistently described both phases was missing, we could not calculate the surface tension in a fully coherent way and therefore treated it as a free parameter. An~alternative approach to computing the finite-size terms is multiple reflection expansion (see, for example,~\cite{Lugones:2015bya,Lugones:2018qgu,Grunfeld:2024ihq}), which we intend to implement in future work.}

In this work, we presented a detailed framework for treating the effects of composition fluctuations and color superconductivity in thermal nucleation theory. Moreover, we showed the impact of such effects on the energy barrier separating the metastable and stable phases.
Due to the high uncertainties in the behavior of strongly interacting matter in the studied regime of baryon density and temperature, our results are model- and parameter-dependent.
A discussion on the astrophysical implications and a quantitative systematic analysis of the impact of the free parameters (namely the~surface tension and the parameters of the SQM EOS) were beyond the scope of this paper, and they will be presented in a work currently in preparation. In~particular, the~theoretical framework presented here will be utilized to investigate the compatibility of the scenario with two families of compact stars  with the early evolution of PNSs.
Moreover, we will test the robustness of our result by using a more sophisticated EOS for the SQM phase. In~particular, we plan to use models for which the vacuum energy (the Bag constant in Bag-like models) and the color superconducting gap $\Delta$ are computed in a thermodynamically consistent way instead of introducing them as free parameters of the model.
Some assumptions used in this work are justified only for systems much larger than the typical droplet size, such as the core of a compact star. However, by~relaxing these assumptions, our general formalism can also be applied to other systems, such as the early universe or the dense and hot nuclear matter produced in heavy-ion~collisions.

\vspace{6pt} 


\authorcontributions{
Conceptualization, M.G., G.P., A.D., and A.L.; methodology, M.G.; validation, M.G. and A.L.; formal analysis, M.G.; investigation, M.G. and G.P.;  writing---original draft preparation, M.G.; writing---review and editing, M.G., G.P, A.D., and A.L.; visualization, M.G.; supervision, G.P, A.D., and A.L. All authors have read and agreed to the published version of the~manuscript.}

\funding{This research received no external~funding.}

\dataavailability{The original contributions presented in this study are included in the article. Further inquiries can be directed to the corresponding author. 
}

\acknowledgments{The authors gratefully acknowledge the CSQCD2024 international conference of the workshop series ``Compact Stars in the QCD phase diagram'' held at the Yukawa Institute for Theoretical Physics in Kyoto for stimulating~discussions.}

\conflictsofinterest{The authors declare no conflicts of~interest.}


\appendixtitles{yes}

\appendixstart
\appendix

\section{Equation of~State}
\label{sec:eos}
A proper QCD-based description of strongly interacting matter is currently not available, since the theory is challenging to solve at finite chemical potentials. Thus, the~more frequent approach is to describe the confined hadronic and deconfined quark phases separately.
Since we are interested in high-energy astrophysical applications other than baryonic degrees of freedom (nucleons, hyperons, delta resonances, up, down, strange quarks, and their antiparticles), we will consider the contributions of leptons (electrons, neutrinos, and~their antiparticles) and thermal bosons (photons and gluons). See~\cite{Oertel_2017,Typel:2022} for reviews on EOSs of high-energy astrophysical systems.
In this section, the EOS models used in the paper will be~described. 
\subsection{Hadrons}
\label{sec:eoshad}

Concerning the hadronic phase, we use a relativistic mean field model in which the strong interaction between the hadrons is mediated by the exchange of massive mesons. The~general form of the Lagrangian density can be written~as follows: \vspace{-12pt}
\begingroup\makeatletter\def\f@size{9}\check@mathfonts
\def\maketag@@@#1{\hbox{\m@th\normalsize\normalfont#1}}
\begin{adjustwidth}{-\extralength}{0cm}
\begin{eqnarray}\label{lagrangian}
{\cal L} &=&
\sum_k\overline{\psi}_k\,[i\,\gamma_{\mu}\,\partial^{\mu}-(M_k-
g_{\sigma k}\,\sigma)
-g_{\omega k}\,\gamma_\mu\,\omega^{\mu}-g_{\phi k}\,\gamma_\mu\,\phi^{\mu}-g_{\rho
k}\,\gamma_{\mu}\,\vec{t} \cdot \vec{\rho}^{\;\mu}]\,\psi_k
+\frac{1}{2}(\partial_{\mu}\sigma\partial^{\mu}\sigma-m_{\sigma}^2\sigma^2)-\,U(\sigma)
\nonumber\\
&& +\frac{1}{2}\,m^2_{\omega}\,\omega_{\mu}\omega^{\mu} +\frac{1}{2}\,m^2_{\phi}\,\phi_{\mu}\phi^{\mu}+\frac{1}{2}\,m^2_{\rho}\,\vec{\rho}_{\mu}\cdot\vec{\rho}^{\;\mu}
+\frac{1}{4}\,c\,(g_{\omega N}^2\,\omega_\mu\omega^\mu)^2
+\frac{1}{4}\,d\,(g_{\rho N}^2\,\vec{\rho}_\mu \cdot \vec{\rho}^\mu)^2
+g_{\rho N}^2A(\sigma,\omega_\mu\omega^\mu) \,  \vec{\rho}_{\mu}\cdot\vec{\rho}^{\;\mu}
\nonumber\\
&&+\overline{\psi}_{\Delta\,\nu}\, [i\gamma_\mu
\partial^\mu -(M_\Delta-g_{\sigma\Delta}
\sigma)-g_{\omega\Delta}\gamma_\mu\omega^\mu
-g_{\rho\Delta}\gamma_\mu I_3 \rho_3^\mu ]\psi_{\Delta}^\nu 
\nonumber\\
&&-\frac{1}{4}F_{\mu\nu}F^{\mu\nu}-\frac{1}{4}P_{\mu\nu}P^{\mu\nu}
-\frac{1}{4}\vec{G}_{\mu\nu}\vec{G}^{\mu\nu}\, ,
\end{eqnarray}
\end{adjustwidth}
\endgroup
where the sum over $k$ runs over the full octet of the lightest baryons ($p$, $n$, $\Lambda$, $\Sigma^+$, $\Sigma^0$, $\Sigma^-$, $\Xi^0$, and $\Xi^-$) interacting with the $\sigma$, $\omega$, $\phi$, and $\rho$ meson fields, $M_k$ is the
vacuum baryon mass of index $k$,  and $U(\sigma)$ is the nonlinear self-iteraction potential of the $\sigma$ meson
\begin{eqnarray}
U(\sigma)=\frac{1}{3}a\,(g_{\sigma
N}\,\sigma)^{3}+\frac{1}{4}\,b\,(g_{\sigma N}\,\sigma^{4}) \,,
\end{eqnarray}
introduced by Boguta and Bodmer~\cite{Boguta:1977xi} to achieve  reasonable compressibility for equilibrium normal nuclear matter. Following~\cite{Steiner:2004fi,Steiner:2012rk}, the~nonlinear function
\begin{eqnarray}
A(\sigma,\omega_\mu\omega^\mu)=\sum_{i=1}^6 a_i\, \sigma_i+\sum_{j=1}^3 b_j \, (\omega_\mu\omega^\mu)^j\,,
\end{eqnarray}
has been introduced in order to fix the experimental range of values of the symmetry energy and the symmetry energy slope $L$. 
In the third line of Equation~(\ref{lagrangian}),  the lagrangian density related to the $\Delta$ isobars ($\Delta^{++},\Delta^{+},\Delta^{0},\Delta^{-}$) is reported, where $\psi_\Delta^\nu$ is the Rarita--Schwinger spinor and $I_3$ is the matrix containing the isospin charges~\cite{Li:1997yh,Lavagno:2010ah}. Finally, the~field strength tensors for the vector mesons are given by the usual expressions
$F_{\mu\nu}\equiv\partial_{\mu}\omega_{\nu}-\partial_{\nu}\omega_{\mu}$,
$P_{\mu\nu}\equiv\partial_{\mu}\phi_{\nu}-\partial_{\nu}\phi_{\mu}$, and
$\vec{G}_{\mu\nu}\equiv\partial_{\mu}\vec{\rho}_{\nu}-\partial_{\nu}\vec{\rho}_{\mu}$.

In the present analysis, we adopted the so-called SFHo parametrization (see~\cite{Steiner:2012rk} for details). The~scalar $\sigma$ meson-hyperon coupling constants have been fitted to reproduce the potential depth of the corresponding hyperon at nuclear matter saturation ~\cite{Schaffner:1993nn,Schaffner:1995th}:
\begin{equation}
U_\Lambda^{N}\!=\!-28 \, {\rm MeV}, \; U_\Sigma^{N}\!=\!+30 \,
{\rm MeV}, \; U_\Xi^{N}\!=\!-18 \, {\rm MeV}\, ,
\end{equation}

Meanwhile, for~the coupling with vector mesons, we used the SU(6) symmetry relations~\cite{Schaffner:1993nn,Fortin_2018}. 

Concerning the meson-$\Delta$ coupling constants, we fixed the ratio $x_{\sigma\Delta}=g_{\sigma\Delta}/g_{\sigma N}=1.0$, $x_{\omega\Delta}=g_{\omega\Delta}/g_{\omega N}=1.0$, and $x_{\rho\Delta}=g_{\rho\Delta}/g_{\rho N}=1.0$. Such a choice resulted in consistency with phenomenological
analysis of the data relative to electron-nucleus coupling, photoabsorption, pion nucleus scattering, and  the experimental flow data of heavy-ion collisions at intermediate energies~\cite{Oset:1987re,Alberico:1994sx,Drago:2014oja,Danielewicz:2002pu}. 

\subsection{Deconfined~Quarks}
\label{sec:eosqua}
In this work, we are interested in two deconfined quark matter phases: an unpaired quark phase and a color-superconducting phase. In~particular, systems smaller (bigger) than the typical diquark pair size will be described with an unpaired (CFL) EOS (see Section~\ref{sec:fwork}). In~both cases, MIT Bag models with perturbative corrections will be extended to finite~temperatures.

\subsubsection{Unpaired~Phase}
\label{sec:eosquaunp}
The unpaired deconfined quark phase will be described using an $\alpha$Bag model (an MIT Bag model with perturbative corrections to the first order) \cite{Fahri_1984,Weissenborn:2011qu} in which the finite temperature generalization is included as explained in~\cite{Fischer_2011}.
The total baryon density, pressure, energy density, entropy density, and free energy density are
\begin{align}
    n_B^{Qunp}(\{\mu_i\},T) &= \frac{1}{3}\sum_{i=u,d,s} n_i^{Q}(\mu_i,T)\\
    P_B^{Qunp}(\{\mu_i\},T) &= \sum_{i=u,d,s} P_i^{Q}(\mu_i,T)-B_{unp}\\
    \varepsilon_B^{Qunp}(\{\mu_i\},T) &= \sum_{i=u,d,s} \varepsilon_i^{Q}(\mu_i,T)+B_{unp}\\
    s_B^{Qunp}(\{\mu_i\},T) &= \sum_{i=u,d,s} s_i^{Q}(\mu_i,T)\\
     f_B^{Qunp}(\{\mu_i\},T) &=  \varepsilon_B^{Qunp}(\{\mu_i\},T)-T\, s_B^{Qunp}(\{\mu_i\},T),
\end{align}
where $n_i^{Q}$ refers to the net number density (namely the difference between the quark and antiquark number densities), while $P_i^{Q}$ and $\varepsilon_i^{Q}$ are the total pressure and energy density (namely the sum between the quark and antiquark quantities) of the quark flavor $i$, respectively, where quark and antiquark are considered to be at the chemical equilibrium with respect to pair production ($\mu_i=-\mu_{\bar{i}}$). 
The contributions of different flavors to the thermodynamical quantities at a finite temperature and chemical potential can be analytically computed in the $\alpha$Bag model only for massless quarks ($m_i=0$) \cite{Fahri_1984,Glendenning2012-ug}
\begin{align}
    n_q(\mu_i,T,m_i=0,\alpha_s)&=\left(\mu_i T^2 +\frac{1}{\pi^2} \mu_i^3\right)\left(1-\frac{2\alpha_s}{\pi}\right)\\
    P_q(\mu_i,T,m_i=0,\alpha_s)&=\frac{7}{60}\pi^2 T^4 \left(1-\frac{
    50 \alpha_s}{21\pi}\right)+\left(\frac{1}{2}T^2 \mu_i^2+\frac{1}{4 \pi^2}\mu_i^4\right)\left(1-\frac{2\alpha_s}{\pi}\right)\\
    \varepsilon_q(\mu_i,T,m_i=0,\alpha_s)&=3P_q(\mu_i,T,m_i=0,\alpha_s)\\
    s_q(\mu_i,T,m_i=0,\alpha_s)&=\frac{7}{15}\pi^2 T^3 \left(1-\frac{
    50 \alpha_s}{21\pi}\right)+T \mu_i^2\left(1-\frac{2\alpha_s}{\pi}\right).
\end{align}
that correspond to lowest-order gluon interaction corrections to the massless Fermi gas with a degeneracy factor $2_{spin}\times 3_{color}=6$.
The thermodynamical quantities at a finite temperature and chemical potential for a gas of non-interacting massive quarks are the Fermi integrals
\begin{align}
    n_q(\mu_i,T,m_i,\alpha_s=0)&=\frac{6}{2 \pi^2}\int_0^{+\infty} k^2\left[\mathbf{f}(k,\mu_i,T,m_i)-\mathbf{f}(k,-\mu_i,T,m_i)\right]dk\\
    P_q(\mu_i,T,m_i,\alpha_s=0)&=\frac{6}{2 \pi^2}\frac{1}{3}\int_0^{+\infty} \frac{k^4}{ E(k,m_i)}\left[\mathbf{f}(k,\mu_i,T,m_i)+\mathbf{f}(k,-\mu_i,T,m_i)\right]dk\\
    \varepsilon_q(\mu_i,T,m_i,\alpha_s=0)&=\frac{6}{2 \pi^2}\int_0^{+\infty} k^2 E(k,m_i)\left[\mathbf{f}(k,\mu_i,T,m_i)+\mathbf{f}(k,-\mu_i,T,m_i)\right]dk\\
     s_q(\mu_i,T,m_i,\alpha_s=0)&=\frac{1}{T}\left[P_q(\mu_i,T,m,0)+\varepsilon_q(\mu_i,T,m,0)-\mu\, n_q(\mu_i,T,m,0)\right],
\end{align}
where the Fermi distribution function is
\begin{equation}
    \mathbf{f}(k,\pm\mu_i,T,m_i) = \frac{1}{e^{\frac{E(k,m_i)\mp \mu}{T}}+1}\label{eq:fermidistr}
\end{equation}
and the single particle energy with a momentum $k$ is
\begin{equation}
    E(k,m_i)=\sqrt{k^2+m^2}.
\end{equation}

In this work, the~Fermi integrals will be computed numerically using the JEL scheme~\cite{Johns_1996} (see also~\cite{Constantinou:2025wxj}). 
Finally, the~quantities for interacting massive quarks will be computed as illustrated in~\cite{Fischer_2011}:
\begin{align}
    n_i^{Q}(\mu_i,T)&=n_q(\mu_i,T,m_i,\alpha_s=0)-\frac{2\alpha_s}{\pi}\left(\mu_i T^2 +\frac{1}{\pi^2} \mu_i^3\right)\label{eq:nEOSQ}\\
    P_i^{Q}(\mu_i,T)&=P_q(\mu_i,T,m_i,\alpha_s=0)-\frac{5\pi \alpha_s}{18} T^4 -\frac{2\alpha_s}{\pi}\left(\frac{1}{2}T^2 \mu_i^2+\frac{1}{4 \pi^2}\mu_i^4\right)\\
   \varepsilon_i^{Q}(\mu_i,T)&= \varepsilon_q(\mu_i,T,m_i,\alpha_s=0)-\frac{15\pi \alpha_s}{18} T^4 -\frac{6\alpha_s}{\pi}\left(\frac{1}{2}T^2 \mu_i^2+\frac{1}{4 \pi^2}\mu_i^4\right)\\
   s_i^{Q}(\mu_i,T)&=s_q(\mu_i,T,m_i,\alpha_s=0)-\frac{
    10 \pi \alpha_s}{9} T^3-\frac{2\alpha_s}{\pi}T \mu_i^2.
\end{align}

The thermodynamical quantities can also be written using $\{n_i\}$ or $n_B,\{Y_i\}$ (where \mbox{$Y_i=n_i/n_B$}) as free variables such that
\begin{equation}
    (\{\mu_i\},T) \rightarrow (\{\mu_i(n_B,Y_i,T)\},T)\rightarrow (n_B,Y_i,T)
\end{equation}
where $\mu_i(n_B,Y_i,T)$ is the chemical potential that solves
\begin{equation}
    n_i^{Q}(\mu_i,T) = n_B\, Y_i.
\end{equation}

\subsubsection{Color-Superconducting Phase (CFL)}
\label{sec:eosquacfl}
The color-superconducting CFL phase will be treated with an extended $\alpha$Bag-inspired model, in~which the contribution of pairing is added to the thermodynamical quantities in a manner similar to that in~\cite{Alford_2005}:
\begin{align}
    n_i^{QCFL}(\{\mu_i\},T) &= n_i^{Q}(\mu_i,T)+\frac{2}{\pi^2}\mu_i \Delta(T)^2\\
    n_B^{QCFL}(\{\mu_i\},T) &= \frac{1}{3}\sum_{i=u,d,s} \left[n_i^{Q}(\mu_i,T)+\frac{2}{\pi^2}\mu_i \Delta(T)^2\right]\\
    P_B^{QCFL}(\{\mu_i\},T) &= \sum_{i=u,d,s} P_i^{Q}(\mu_i,T)+\frac{1}{\pi^2}\Delta(T)^2\sum_{i=u,d,s}\mu_i^2-B_{CFL}\\
     \varepsilon_B^{QCFL}(\{\mu_i\},T) &= f_B^{QCFL}(\{\mu_i\},T)+T\,  s_B^{QCFL}(\{\mu_i\},T)        \\
    s_B^{QCFL}(\{\mu_i\},T) &= \sum_{i=u,d,s} s_i^{Q}(\mu_i,T)+\frac{2}{\pi^2}\Delta(T)\frac{\partial \Delta(T)}{\partial T}\sum_{i=u,d,s}\mu_i^2\\
   f_B^{QCFL}(\{\mu_i\},T) &= -P_B^{QCFL}(\{\mu_i\},T)+\sum_i  \mu_i n_i^{QCFL}(\{\mu_i\},T),
\end{align}
where $\Delta(T)$ is a finite temperature extension of the gap parameter~\cite{Schmitt_2010,Alford:2007xm}
\begin{equation}
    \Delta(T)=\Theta(T_c-T)\Delta_0\sqrt{1-\frac{T}{T_c}},
\end{equation}
where $\Delta_0$ is the gap parameter at zero temperature and~$T_c\simeq 2^{1/3} 0.57 \Delta_0$ \cite{Schmitt_2010,Alford:2007xm}. Unlike the unpaired case, in the~CFL phase, the~chemical potentials $\{\mu_i\}$ are not free variables but are fixed by the conditions that the number densities of the up, down and strange quarks are equal. Specifically, we have
\begin{equation}
  (\{\mu_i\},T) \rightarrow (\{\mu_i(n_B,T)\},T) \rightarrow (n_B,T) \,\,\,\,\,\,\text{ $i=u,d,s$}
\end{equation}
using
\begin{align}
    n_u^{QCFL}(\{\mu_i\},T)&= n_d^{QCFL}(\{\mu_i\},T)= n_s^{QCFL}(\{\mu_i\},T)
\end{align}

\subsection{Leptons}
\label{sec:eoslep}
Electrons and positrons will be treated as massive non-interacting Fermi gas with a degeneracy of two:
\begin{align}
    n_e(\mu_e,T)&=\frac{2}{2 \pi^2}\int_0^{+\infty} k^2\left[\mathbf{f}(k,\mu_e,T,m_e)-\mathbf{f}(k,-\mu_e,T,m_e)\right]dk\\
    P_e(\mu_e,T)&=\frac{2}{2 \pi^2}\frac{1}{3}\int_0^{+\infty} \frac{k^4}{ E(k,m_e)}\left[\mathbf{f}(k,\mu_e,T,m_e)+\mathbf{f}(k,-\mu_e,T,m_e)\right]dk\\
    \varepsilon_e(\mu_e,T)&=\frac{2}{2 \pi^2}\int_0^{+\infty} k^2 E(k,m_e)\left[\mathbf{f}(k,\mu_e,T,m_e)+\mathbf{f}(k,-\mu_e,T,m_e)\right]dk\\
     s_e(\mu_e,T)&=\frac{1}{T}\left[P_e(\mu_e,T)+\varepsilon_e(\mu_e,T)-\mu_e\, n_e(\mu_e,T)\right].
\end{align}

\noindent Again, the~Fermi integrals will be computed numerically using the JEL scheme~\cite{Johns_1996}.
Neutrinos will be treated as massless non-interacting Fermi gas with a degeneracy of one:
\begin{align}
    n_{\nu}(\mu_{\nu},T)&=\frac{1}{6}\left(\mu_{\nu} T^2 +\frac{1}{\pi^2} \mu_{\nu}^3\right)\\
    P_{\nu}(\mu_{\nu},T)&=\frac{1}{6}\left(\frac{7}{60}\pi^2 T^4 +\frac{1}{2}T^2 \mu_{\nu}^2+\frac{1}{4 \pi^2}\mu_{\nu}^4\right)\\
    \varepsilon_{\nu}(\mu_{\nu},T,)&=3P_{\nu}(\mu_{\nu},T)\\
    s_{\nu}(\mu_{\nu},T)&=\frac{1}{6}\left(\frac{7}{15}\pi^2 T^3 +T \mu_{\nu}^2\right),
\end{align}
As for the quark phase, the~thermodynamical quantities of electrons (neutrinos) can be written in terms of $n_e,T$ ($n_{\nu},T$) as independent variables.  
Muons could also, in~principle, play a role in high-energy astrophysical systems (see, for example,~\cite{Loffredo:2022prq}), at~least in the hadronic phase. However, in~this work, they were neglected.

\subsection{Bosons}
\label{sec:eosbos}
Photons will be treated as an ideal boson gas with zero chemical potential (i.e., as  blackbody radiation):
\begin{align}
P_{\gamma}(T)&=\frac{\pi^2}{45}T^4\\
\varepsilon_{\gamma}(T)&=3P_{\gamma}(T)\\
s_{\gamma}(T)&=\frac{4\pi^2}{45}T^3.
\end{align}

The contribution of thermal gluons is~\cite{Glendenning2012-ug}
\vspace{3pt}
\begin{align}
    P_g(T)&=\frac{8 \pi^2}{45}T^4\left(1-\frac{15 \alpha_s}{4 \pi}\right)\\
    \varepsilon_g(T)&=3 P_g(T)\\
    s_g(T)&=\frac{32 \pi^2}{45}T^3\left(1-\frac{15 \alpha_s}{4 \pi}\right).
\end{align}

\subsection{Total~EOS}
The total thermodynamical quantities in the hadronic and quark phases are
\begin{adjustwidth}{-\extralength}{0cm}
\begin{align}
    X_H(n_B,\{Y_i\},T)&= X_B^H(n_B,\{Y_h\},T)+ X_e(n_B\,Y_e,T)+ X_{\nu}(n_B\,Y_{\nu_e},T)+2X_{\nu}(0,T)+X_{\gamma}(T)\\
    X_{Qunp}(n_B,\{Y_i\},T)&= X_B^{Qunp}(n_B,\{Y_q\},T)+ X_e(n_B\,Y_e,T)+ X_{\nu}(n_B\,Y_{\nu_e},T)+2X_{\nu}(0,T)+X_{\gamma}(T)+X_{g}(T)\\
    X_{QCFL}(n_B,T)&= X_B^{QCFL}(n_B,T)+ X_{\nu}(n_B\,Y_{\nu_e},T)+2X_{\nu}(0,T)+X_{\gamma}(T)+X_{g}(T),
    \end{align}
    \end{adjustwidth}
where $X=P,\varepsilon,s,f$. 
In principle, the~number densities $n_B,\{Y_i\}$ (or the chemical potentials $\{\mu_i\}$) are free variables. However, they can be fixed using some conditions of the system. For~example, if~an unpaired quark phase is in weak equilibrium and charge-neutral, then the five~relations
\begin{align}
    \mu_u(n_B,\{Y_i\},T)+\mu_e(n_B,\{Y_i\},T)&=\mu_d(n_B,\{Y_i\},T)+\mu_{\nu_e}(n_B,\{Y_i\},T)\\
    \mu_d(n_B,\{Y_i\},T)&=\mu_s(n_B,\{Y_i\},T)\\
    1&=\frac{1}{3}Y_u+\frac{1}{3}Y_d+\frac{1}{3}Y_s\\
    0&=\frac{2}{3}Y_u-\frac{1}{3}Y_d-\frac{1}{3}Y_s-Y_e\\
    Y_L&=Y_e+Y_{\nu_e},
\end{align}
can be used to fix $Y_u,Y_d,Y_s,Y_e,Y_{\nu_e}$, and $n_B$, $Y_L$, and $T$ remain  independent~variables.

In this work, we  fix the values at $\alpha_s =\pi/2 \times 0.1$, $m_s=100$ MeV, $m_u=m_d=0$, $B_{unp}^{1/4}=175$~MeV, $B_{unp}^{1/4}=135$ MeV, and $\Delta_0=80$ MeV.
These parameters were chosen in order to have a CFL (unpaired) phase that fullfilled (did not fullfill) the Witten hypothesis on the absolute stability of SQM~\cite{Witten:1984rs,Bodmer:1971we}. 
In particular, the~CFL parametrization is similar to the one used in~\cite{Bombaci:2020vgw} in the context of the two-family scenario of compact stars, while the Bag of the unpaired phase is to the same order of the one used in works using unpaired deconfined quark matter (see, for example,~\cite{Constantinou:2021hba,Constantinou:2023ged,Constantinou:2025wxj,Guerrini:2024gzu,Fischer_2011}).
A more detailed discussion on the astrophysical implications will be discussed in a future work.  
Some possible explanations for a different Bag for the unpaired and color-superconducting phases will be investigated in future works by computing the Bag and the gap for the two phases consistently without~considering them free variables.
For example, different $B$ values for different $\Delta_0$ values were obtained in~\cite{Blaschke:2022egm,Gartlein:2023vif}, where a color-superconducting phase was computed with an NJL-like model and~then fitted to a Bag-like model~\cite{Alford_2005}.



\printendnotes[custom] 

\reftitle{References}



\PublishersNote{}

\end{document}